\documentclass[journal]{IEEEtran}
\IEEEoverridecommandlockouts
\usepackage{cite}
\usepackage{tabularx}
\usepackage{amsmath,amssymb,amsfonts}
\usepackage{algorithmic}
\usepackage{multirow}
\usepackage{ragged2e,array,booktabs}
\usepackage{graphicx}
\usepackage{textcomp}
\usepackage{xcolor}
\usepackage{textgreek}

\usepackage{bookmark}
\usepackage{cuted}
\usepackage{comment}
  \pdfoutput=1

\usepackage{stfloats}

\usepackage{subfigure}
\usepackage{float} 
\def\BibTeX{{\rm B\kern-.05em{\sc i\kern-.025em b}\kern-.08em
		T\kern-.1667em\lower.7ex\hbox{E}\kern-.125emX}}
\usepackage{tikz}
\usetikzlibrary{arrows}
\tikzstyle{int}=[draw, fill=white!20, minimum width=3cm, minimum height=1cm]
\newcommand{\etal}{\textit{et al}. }
 
\begin{document}
\title{The End-to-End Molecular Communication Model of Extracellular Vesicle-based Drug Delivery}
\author{Hamid Khoshfekr Rudsari, Mohammad Zoofaghari, Mladen Veleti\'{c},~\IEEEmembership{Member,~IEEE}, Jacob Bergsland, Ilangko Balasingham,~\IEEEmembership{Senior Member,~IEEE}
	\thanks{This work was supported in part by the Research Council of Norway (RCN: CIRCLE Communication Theoretical Foundation of Wireless Nanonetworks) under Grant 287112.}
	\thanks{This article was presented in part at the ACM Conference on Embedded Networked Sensor Systems, Coimbra, Portugal, November 2021.}
	\thanks{Hamid Khoshfekr Rudsari is with the Institute of Clinical Medicine, Faculty of Medicine, University of Oslo, 0316 Oslo, Norway, and the Intervention Centre, Oslo University Hospital, 0372 Oslo, Norway. e-mail: h.k.rudsari@studmed.uio.no} 
	\thanks{Mohammad Zoofaghari is with the	Department of Electrical Engineering, Yazd University, Yazd 89195-741, Iran, and the Intervention Centre, Oslo University Hospital, 0372 Oslo, Norway. e-mail: zoofaghari@yazd.ac.ir}
	\thanks{Mladen Veleti\'{c} and Ilangko Balasingham are with the Department of Electronic Systems, Norwegian University of Science and Technology, 7491 Trondheim, Norway, and the Intervention Centre, Oslo University Hospital, 0372 Oslo, Norway. e-mail:\{mladen.veletic@ntnu.no, ilangko.balasingham@ntnu.no\}}
	\thanks{Jacob Bergsland is with the Intervention Centre, Oslo University Hospital, 0372 Oslo, Norway. e-mail: Jacob.bergsland@ous-hf.no}
}
\maketitle

\begin{abstract}
A closer look at nature has recently brought more interest in exploring and utilizing intra-body communication networks composed of cells as intrinsic, perfectly biocompatible infrastructures to deliver therapeutics. Naturally occurring cell-to-cell communication systems are being manipulated to release, navigate, and take up soluble cell-derived messengers that are either therapeutic by nature or carry therapeutic molecular cargo in their structures. One example of such structures is extracellular vesicles (EVs) which have been recently proven to have favorable pharmacokinetic properties, opening new avenues for developing the next generation biotherapeutics. In this paper, we study theoretical aspects of the EV transfer within heart tissue as a case study by utilizing an information and communication technology-like approach in analyzing molecular communication systems. Our modeling implies the abstraction of the EV releasing cells as transmitters, the extracellular matrix as the channel, and the EV receiving cells as receivers. Our results, derived from the developed analytical models, indicate that the release can be modulated using external forces such as  electrical signals, and the transfer and reception can be affected by the extracellular matrix and plasma membrane properties, respectively. The results can predict the EV biodistributions and contribute to avoiding unplanned administration, often resulting in side- and adverse effects.
\end{abstract}



\section{Introduction}
\IEEEPARstart{E}{xtracellular} vesicles (EVs) are natural, cell-derived messengers that act as vehicles of different biomolecules such as proteins, nucleic acids, and lipids, thus serving as potential candidates for treating different disorders~\cite{tetta2013extracellular}. EVs can be engineered for carrying drugs and targeting diseased cells under the control of targeted drug delivery systems for different types of disorders \cite{ELSHARKASY2020332,herrmann2021extracellular}. 
However, unplanned, systematic administration of nanoparticles and EVs may accumulate in sites beyond the tissues of therapeutic interest, resulting in off-target and adverse effects. Therefore, a required step towards optimizing therapeutic efficacy is to study their propagation and biodistribution after the administration. Such studies comprise both theoretical and experimental approaches; the former are the subject of our interest. 


The propagation and transport of different types of micro-and nanoparticles in the human body have been theoretically studied. Syková \etal mathematically modeled the diffusion of molecules/drugs in the brain extracellular space \cite{doi:10.1152/physrev.00027.2007}. The diffusion was derived utilizing structural descriptions, including the volume fraction and tortuosity, which are the physical parameters discussed later in this paper whose values were derived from experiments. The diffusion model was simplified under the spherically symmetric coordinate system without considering the degradation of molecules/drugs by different factors, such as enzymatic reactions and binding to various receptors in the brain extracellular space or degradation due to the natural half-life of the considered molecules \cite{nicholson1995interaction}. Leedale \etal performed spatiotemporal modeling of the drug transport in liver spheroids \cite{doi:10.1098/rsfs.2019.0041}, by importing the geometry of hepatocytes into a Voronoi diagram comprised of partitions, each of which introduces one cell with nuclei. Although the model considered the metabolism of drugs in the hepatocytes as part of the degradation, it was simplified by assuming that the diffusion problem was radially symmetric. Also, the volume fraction as part of the organ's structural properties was not considered. Furthermore, the release rate in this model was constant while its dynamics could change the results significantly~\cite{9488662}. Mok \etal modeled the spatiotemporal propagation of the herpes simplex virus in the interstitial tumor micro-environment \cite{mok2009mathematical}. They also specified boundary conditions based on the assumed radial symmetry of the tumor micro-environment, and considered a constant virus release rate according to intratumor injection method. They also modeled the internalization of viruses into the tumor cells with no limit on the number of internalized particles. The proposed model can be enhanced by limiting the number of available bounding sites, which changes the number of internalized particles.
Finally, Ebrahimi \etal studied effects of the bloodstream and drug carrier types in a vesicle-based drug delivery system in patient-specific geometry of abdominal aortic aneurysm \cite{ebrahimi2021drug}. Their model was based on using computed tomography (CT) scan and they evaluated the surface density of adhered nanoparticles to inner wall of abdominal aortic aneurysm. 

\begin{figure*}[!] 
	\centering
	\includegraphics[width=\linewidth]{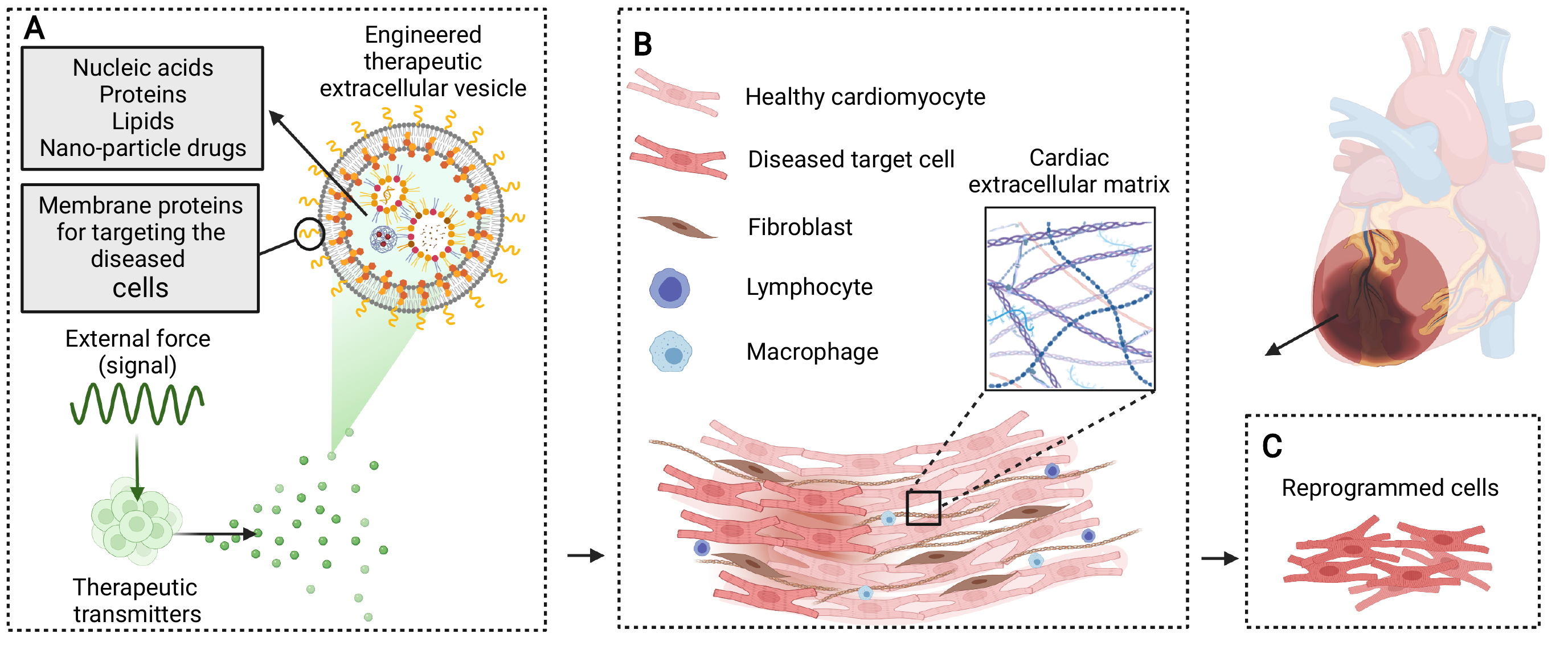}	
	\vspace{-0.5 cm}
	\caption{A graphical representation of the proposed end-to-end drug delivery for cardiovascular disease. A) The therapeutic transmitters as human induced pluripotent stem cells (HiPSCs) release engineered therapeutic extracellular vesicles (EVs) based on the control signal from an external device which modulates the EV release rates by changing the cytosolic Ca$^\text{2+}$ levels. B) The cardiac space is comprised of different cell types such as cardiomyocytes, fibroblasts, lymphocytes and macrophages \cite{doi:10.1016/j.jacc.2020.03.024}. The extracellular matrix (ECM) in the cardiac tissue has an important role in cellular and architectural functions and is made up of filamentous proteins and different types of collagens such as collagen type-I and type-III. C) The sick target cells can be reprogrammed by internalizing the therapeutic EVs. The illustration is	created using BioRender.com.}
	\label{fig:ECM}
\end{figure*}

In this paper, we aim to study the EV biodistribution. The heart is the selected organ for our analyses, wherein EVs indicate having positive therapeutic effects on cardiovascular diseases (CVDs) \cite{de2020native,10.1093/eurheartj/ehw304}. The presented available computational and analytical methods, cannot be readily used in the EV biodistibution analyses because they are unable to address the critical challenges in the modeling of the propagation and transport of micro-and nanoparticles in the cardiac extracellular space, which is hindered by complex interstitial matrix between cells. A potential strategy to avoid this issue is to use the molecular communications (MC) paradigm \cite{akyildiz2008nanonetworks,6208883,arjmandi2021extracellular}, which utilizes mathematical tools widely applied in communications engineering to provide a systems approach for measuring information exchange in biological communication networks. The MC paradigm has been utilized in proposing an initial pharmacokinetic model of therapeutic nanoparticles’ propagation considering advection and diffusion in the blood vessel \cite{7103028}. 
We also proposed a MC-based drug delivery system comprising engineered human induced pluripotent stem cells (HiPSCs) differentiated into ventricular \cite{9488662} and atrial cardiomyocytes~\cite{10.1145/3477206.3477479}. 
We now present the MC-based analysis of an end-to-end drug delivery system comprising engineered HiPSC-cardiomyocytes, the cardiac extracellular space and sick cardiac cells. Specifically, we build the spatiotemporal model considering i) the \textit{release} of EVs from HiPSC-derived cardiomyocytes, ii) the \textit{propagation} of EVs in a 3-dimensional (3D) cardiac extracellular space, and iii) the \textit{internalization} of EVs into the sick cardiac cells through different reception processes. A graphical representation of the considered system for the treatment of CVDs is presented in Fig.~\ref{fig:ECM}. 

The contributions of this paper can be summarized as follows:
\begin{enumerate}
	\item 
	We model the stochastic EV release, which has not been studied before. The modeling is inspired by our recent work \cite{9488662}, and is now extended by considering	a Poisson process to assess the distribution of release	events from various cellular compartments in predefined time frames.   
	
	\item We model the propagation of EVs in the cardiac extracellular matrix (ECM) by considering the unique properties of ECM -- \textit{tortuosity} and \textit{volume fraction} -- that describe the hindrance sources slowing down the free diffusion and propagation of EVs. The spatiotemporal modeling is formulated based on an advection-diffusion problem, including a 3D partial differential equation (PDE) and considering the EV release rate as a source. We model the injection of EVs into the ECM by a Gaussian function, unlike most works that consider a point source for the release scheme. We also consider degradation of EVs in the ECM as part of their natural half-life, extracellular binding to non-target cells and the advection of EVs \cite{CHAROENVIRIYAKUL2017316,buzas2018molecular}, without considering any symmetrical simplifications.
	
	\item We derive an analytical solution of the 3D-PDE model for EV propagation based on a Green's function. The proposed analytical solution is verified through numerical analysis using the finite element method (FEM). The proposed analytical solution is the first analytical solution for the advection-diffusion of EVs by considering unique ECM properties and a spatiotemporal modeling for the injection function of the release of EVs into the cardiac ECM. The analytical solution can be applied to other tissues by updating the required parameters.
	
	\item We model the internalization of EVs using systems of ordinary differential equations (ODEs) with two reception mechanisms, i.e., ligand-receptor interaction and endocytosis, which are the primary means of internalization and docking to the target cells \cite{FRENCH201748}.
\end{enumerate}


The rest of the paper is organized as follows. In Section \ref{sec:biological_model}, we explain the biological background information necessary for understanding the analyzed mechanisms in the considered end-to-end EV-based drug delivery system. In Section \ref{sec:drug_delivery}, we study the stochastic process behind the release events of EVs. In Section \ref{sec:ECM}, we model the propagation of EVs by proposing a 3D advection-diffusion PDE. In this section, we also derive an analytical solution to the 3D-PDE model by the Green's function. In Section \ref{sec:EV_Internalization}, we model the internalization of EVs through endocytosis mechanisms. In Section \ref{sec:results}, we provide the numerical analysis of the spatiotemporal modeling of the end-to-end drug delivery system and conclude the paper in Section \ref{sec:conclusion}.

\section{Biological Background} \label{sec:biological_model}
In this section, we explain the biological aspects necessary to understand the considered end-to-end drug delivery system.

\subsection{Therapeutic Transmitters} \label{subsec:biological_tx}
The therapeutic transmitters shown in Fig.~\ref{fig:ECM}-A are assumed as HiPSCs. An external device controls the release rate of modulated EVs from HiPSC-derived cardiomyocytes by manipulating the calcium ion (Ca$^\text{2+}$) influx \cite{9488662}. Modulating the cytosolic Ca$^\text{2+}$ levels in the HiPSC-derived cardiomyocytes lead to changing the positions of phospholipids in the cells' membrane \cite{sluijter2014microvesicles}, and finally result in a controlled manner of EV release. Exocytosis in the engineered HiPSC-derived cardiomyocytes, the biological transport of EVs from inside to outside of the cell membrane, occurs in two main microdomains as submembrane and L-type Ca$^\text{2+}$ channels (LTCCs). This is because the exocytosis sites are close to the LTCCs, and Ca$^\text{2+}$-mediated exocytosis happens in these nanodomains \cite{gilbert2020calcium,BUCURENCIU2008536,Beaumont10700}. 

One example of controlling cellular functions in cells is using novel methods in nanotechnology such as employing magnetic nanoparticles at the voltage-gated Ca$^\text{2+}$ channels \cite{dobson2008remote}. The nanoparticles can bind to the surface of the cells and control Ca$^\text{2+}$ influx from Ca$^\text{2+}$ channels such as transient receptor potential (TRP) channels into the cells to manipulate the cytosolic (the internal environment of the cell embedding different organelles and sub-cellular compartments) Ca$^\text{2+}$ levels \cite{https://doi.org/10.1111/bph.14792,WU2021101187}. It is experimentally demonstrated that energy-based stimulation as ultrasound \cite{castellanos2016therapeutic} and localized magnetic fields \cite{nag2020nanoparticle} can control nanoparticle-based manipulation of cellular functions. 


\subsection{Cardiac Extracellular Matrix (ECM)}
The extracellular space of a tissue is made up of ECM components which are non-cellular compartments of the tissue actively involved in the cellular functions. The ECM also has regulatory roles in establishing architectural tissue functions \cite{doi:10.1016/j.jacc.2020.03.024}. Cardiac ECM affects cell migration, tissue growth, fibrosis, progenitor cell self-renewal, and morphogenesis, which are active and dynamic regulatory functions \cite{doi:10.1126/science.1191035,poh2014generation,ENGLER2006677}. ECM is formed with filamentous proteins, chains of proteins bundled together to increase strength and rigidity. ECM is also formed with other proteins such as proteoglycans and long linear polysaccharides such as glycosaminoglycans. Different types of collagens are found in cardiac ECM where collagen type I contributes to tensile strength, which is resistant to the length-wise stress, and collagen type III contributes toward the elastic features of cardiac ECM \cite{10.1093/cvr/17.1.15}. We represent cardiac ECM in our study in Fig.~\ref{fig:ECM} where different components of the ECM are shown.

EVs can interact with ECM components such as matrix molecules. The interaction of EVs and the ECM components---\textit{extracellular binding}---is mainly mediated by the integrins (a type of receptor proteins) on the surface of EVs \cite{buzas2018molecular}.  One of the complexes that leads to extracellular binding is the fibronectin-integrin complexes \cite{sung2015directional}. Fibronectins are matrix molecules and an insoluble network that play an essential role in organizing the tissue structure. It is believed that the inhibition of specific integrins can result in less binding of EVs to ECM and specific cells, which has been proposed as a novel type of cancer treatment \cite{altei2020inhibition}.

Another feature that results in degradation of EVs in the ECM is their half-life. EVs have a half-life between 2~min to 30~min depending on the three factors such as the location where they are administered, their cell-type origins, and the presence of target cells for their internalization \cite{CHAROENVIRIYAKUL2017316}.

\subsection{EV Internalization}
EVs interact with recipient cells by targeting and docking to the cells' surface. It is believed that EVs can precisely target specific cells through their membrane proteins while crossing biological barriers such as the blood brain barrier \cite{RANA20121574,GUDBERGSSON2019108}. We show a representation of EVs membrane proteins in Fig.~\ref{fig:ECM}. Once EVs attach to the target cell's surface, they either interact with the cell using the membrane proteins through ligand-receptor interactions and activate the cell's receptor \cite{maas2017extracellular,vinas2018receptor,10.3389/fcell.2018.00018} or undergo the process called endocytosis \cite{COSSETTI2014193,FRENCH201748}.

EVs stimulate cell signaling pathways by their membrane proteins and target receptors on the plasma membrane surface of the cells. This type of interaction enables EVs to \textit{address} specific cells where it is possible to engineer the EVs' membrane proteins to address the target cells \cite{costa2015pancreatic}. The ligand-receptor interaction of EVs can lead to internalization of their cargo or other biological effects that are originated from EV-transported growth factors, ECM proteins, and angiogenic factors \cite{maas2017extracellular}.

EV uptake can happen by an endocytic mechanism at the target cell. This mechanism begins with forming a pit resulting from membrane invagination, and is then coated with a protein called clathrin. An enzyme family in the target cell called dynamin un-coat the clathrin-coated pits, which results in releasing the pits to the cytosolic environment of the cell and finally leads to internalization of EVs \cite{mcmahon2011molecular}. Other types of endocytosis, such as caveolin-mediated and lipid raft-mediated, as well as phagocytosis and macropinocytosis, are other possible mechanisms of EV's internalization which are out of the scope of this paper \cite{FRENCH201748}.

\section{Extracellular Vesicle Release Modeling} \label{sec:drug_delivery}
The Poisson process is widely used to model counting processes that have specific rates for event occurrences, but in uncertain timing of events \cite{papoulis1989probability}. For the stochastic modeling, we consider small EVs (often referred to as exosomes) as a type of EVs whose release events presumably occur in time intervals $(t, t+\Delta t]$ in a Poisson process with a rate of
\begin{align}
	\phi(t) = \dfrac{\gamma(t)}{\mathbb{E}[C_\text{MVB}]}, \label{eq:occurance_rate}
\end{align}
where $\gamma(t)$ is the EV release rate and $\mathbb{E}[C_\text{MVB}]$ is the average of concentration of EVs in multivesicular bodies (MVBs) which are cargoes of EVs that fuse to the cell membrane and ultimately result in the release of EVs. We previously derived the EV release rates from the submembrane and LTCC microdomains respectively as \cite{9488662} 
\begin{subequations}
	\begin{align}
		\begin{split}
			&\gamma_\text{s}(t) = \dfrac{[\text{Ca}^{\text{2+}}]^{\ell_n}_\text{s}}{[\text{Ca}^{\text{2+}}]^{\ell_n}_\text{s} + M_n^{\ell_n}}, \label{eq:rate_submembrane}
		\end{split}
		\\
		\begin{split}
			&\gamma_\text{LTCC}(t) = A_v I_v I_c \dfrac{[\text{Ca}^{\text{2+}}]^{\ell_m}_\text{open}}{[\text{Ca}^{\text{2+}}]^{\ell_m}_\text{open} + M_m^{\ell_m}} \\&~~~~~~~~~~~~~+ (1 -A_v I_v I_c) \dfrac{[\text{Ca}^{\text{2+}}]^{\ell_m}_\text{close}}{[\text{Ca}^{\text{2+}}]^{\ell_m}_\text{close} + M_m^{\ell_m}}, \label{eq:rate_ltcc}
		\end{split}
	\end{align}
\end{subequations}
where $[\text{Ca}^{\text{2+}}]_\text{s}$, $[\text{Ca}^{\text{2+}}]_\text{open}$, and $[\text{Ca}^{\text{2+}}]_\text{close}$ are the Ca$^\text{2+}$ concentration in the submembrane space, the Ca$^\text{2+}$ concentration in LTCC microdomain in case of opening and closing the LTCCs, respectively. $A_v$, $I_v$, and $I_c$ are the gating variables in the Ca$^\text{2+}$ dynamics modeling, and $\ell_n$, $M_n$, $\ell_m$, and $M_m$ are parameters for exocytosis modeling \cite{montefusco2015mathematical}. The cumulative release rate is defined as $\gamma(t) = \gamma_\text{s}(t) + \gamma_\text{LTCC}(t)$.

By considering an average of $\mathbb{E}[N_\text{MVB}]$ EVs in the membrane-enclosed lumen of each MVB and assuming their spherical shape with a diameter of $d_\text{MVB}$, following the molar concentration rule \cite{shang2014nanoparticle}, the average concentration of EVs inside a MVB is
\begin{align}
	\mathbb{E}[C_\text{MVB}] = \dfrac{6 \mathbb{E}[N_\text{MVB}]}{\pi N_\text{A}  d^3_\text{MVB}}, \label{eq:c_mvb}
\end{align}
where $N_\text{A}$ is the Avogadro constant. The number of release events denoted by $k$ in the Poisson process then follows a Poisson distribution in each time interval $(t, t+\Delta t]$ as 
\begin{align}
	k \sim \mathcal{P} \left(\phi(t) \cdot \Delta t\right). \label{eq:poisson}
\end{align}

\section{Extracellular Vesicle Propagation Modeling} \label{sec:ECM}
In what follows, we propose the spatiotemporal modeling of the cardiac ECM considered as the EV propagation medium/channel. 

\subsection{3D-PDE Problem for the Diffusion and Advection of EVs}

The advection-diffusion PDE problem for the propagation of EVs in cardiac ECM, with the boundary condition (BC) and the initial condition (IC), is
\begin{subequations} \label{eq:diff_conv_all}
			\begin{align}
		&\frac{\partial C\left(\mathbf{x},t\right)}{\partial t} = \nabla \cdot \left(\mathbf{K} \cdot \vec{\nabla} C\left(\mathbf{x},t\right)\right) - \mathbf{v} \cdot \vec{\nabla} C\left(\mathbf{x},t\right) \nonumber \\& \hspace{2cm} - P(t) + \Gamma(\mathbf{x}, t, \mathbf{X_0}), \quad \text{in}~~\Omega \times T \label{eq:PDE}
		\\
		&\text{BC:}~\vec{n}\cdot\left(\textbf{K} \vec{\nabla} C\left(\mathbf{x},t\right)\right) = 0, \hspace{1.65cm}	
		 \text{on}~~\partial\Omega_N  \times T \label{eq:BC}
		\\
		&\text{IC:}~C(\mathbf{x}, t_L) = 0, \hspace{3.1cm} \text{in}~~ \Omega \label{eq:IC}
	\end{align}
\end{subequations}
where $\mathbf{x} = (x, y, z)$ is a point in the spatial domain $\Omega$, wherein the boundary is denoted as $\partial \Omega_N$,  $T = (t_L, t_R)$ is the temporal domain and $\vec{n}$ is the outward unit normal. We study \eqref{eq:PDE}-\eqref{eq:IC} in what follows.

\subsubsection{PDE \eqref{eq:PDE}}
Eq. \eqref{eq:PDE} consists of four parts. The first part models the concentration dynamics over the diffusion by taking the divergence of the gradient of the EV concentration where the diffusivity tensor $\mathbf{K}$ models the diffusion in $x$-, $y$-, and $z$-direction. Due to an anisotropic nature of the cardiac ECM \cite{stoppel2015anisotropic}, we consider different values of tortuosity in the three directions; tortuosity denoted by $\lambda$ models how a convoluted pathway of a porous medium differs from an obstacle-free medium. Then, the effective diffusion coefficient in the medium is $D^* = D/\lambda^2$ \cite{doi:10.1152/physrev.00027.2007}. By considering the cardiac ECM as a non-isotropic micro-environment, the diffusivity tensor $\mathbf{K}$ is
\begin{align} \label{eq:diffusivity_tensor}
	\mathbf{K} = \begin{bmatrix}
		\dfrac{D}{\lambda_x^2} & \dfrac{D}{\lambda_y^2} & \dfrac{D}{\lambda_z^2}
	\end{bmatrix} \times \mathbf{I}, 
\end{align}
where $D$ is the diffusion coefficient of EVs in the cardiac ECM and $\lambda_x$, $\lambda_y$, and $\lambda_z$ are respectively the tortuosity of the ECM in the three directions. Also, $\mathbf{I}$ in \eqref{eq:diffusivity_tensor} is the 3 $\times$ 3 identity matrix.

The second term in \eqref{eq:PDE} models the concentration dynamics over EV advection. Due to viscoelastic properties of cardiac ECM \cite{jansen2017mechanotransduction}, the EVs can affected by advection. In this regard, $\mathbf{v} = (v_x, v_y, v_z)$ is the velocity of ECM in the three directions. The velocity profile is multiplied by the gradient of the EV concentration, thus changing the gradient vector's elements in the three directions. 

The third term of \eqref{eq:PDE} formulates the degradation of EVs over their half-life and extracellular binding to the ECM. We consider an exponential degradation over time to account for the half-life of EVs and model the extracellular binding to the ECM as a first-order degradation reaction. The overall degradation of EVs is modeled as
\begin{subequations} \label{eq:degradation}
	\begin{align}
		P(t) &= \underbrace{\frac{ C\left(\mathbf{x},t\right)}{\alpha} \left(1 - \exp\left(\frac{-\left(t - t_L\right) }{\sigma}\right)  \right)}_{\text{Half-life}} + \underbrace{k_e \frac{ C\left(\mathbf{x},t\right)}{\alpha}}_{\text{Extracellular binding}}, \label{eq:degradation-exp}
		\\
		\sigma &= \frac{\Lambda_{1/2}}{\ln(2)}, \label{eq:degradation-sigma}
	\end{align}
\end{subequations}
where $\Lambda_{1/2}$ is the half-life of EVs and $\alpha = V_\text{ECM}/V_\text{Total}$ is the volume fraction of the ECM, where $V_\text{ECM}$ is the volume of ECM and $V_\text{Total}$ is the total volume of heart tissue. Volume fraction models the relative volume accessible to EVs compared to the total volume of the cardiac space \cite{doi:10.1152/physrev.00027.2007}. $\sigma$ in \eqref{eq:degradation-sigma} is the decay rate for the degradation of EVs with respect to the half-life $\Lambda_{1/2}$. Also, $k_e$ is the extracellular binding rate of the EVs.  

The fourth term in \eqref{eq:PDE} models the source of EVs injected into the ECM. In the proposed end-to-end drug delivery model, the therapeutic transmitters release EVs from $\mathbf{X_0} = (x_L, y_L, z_L)$ point in the ECM micro-environment. We propose to consider a Gaussian function for the EVs injection into the ECM based on the release rate $\gamma(t)$ as
\begin{align}
	\Gamma(\mathbf{x}, t, \mathbf{X_0}) &= \frac{\gamma(t)}{\alpha} \exp\bigg( \frac{- (x-x_L)^2}{2\sigma^2_x} + \frac{- (y-y_L)^2}{2\sigma^2_y} \nonumber \\& \quad+ \frac{- (z-z_L)^2}{2\sigma^2_z} \bigg), \label{eq:release}
\end{align}
where $\sigma_i$ for $i \in \{x, y, z\}$ can specify the injection function of EVs from the transmitters. Then, the 3D-PDE for the diffusion-advection of EVs in \eqref{eq:PDE} becomes
		\begin{align} \label{eq:diff_conv_second}
			\frac{\partial C\left(\mathbf{x},t\right)}{\partial t} &= \nabla \cdot \left(\mathbf{K} \cdot \vec{\nabla} C\left(\mathbf{x},t\right)\right) - \mathbf{v} \cdot \vec{\nabla} C\left(\mathbf{x},t\right) \nonumber \\&\quad+ \frac{ C\left(\mathbf{x},t\right)}{\alpha} \left(\exp\left(\frac{t_L - t }{\sigma}\right) - 1 - k_e\right) \nonumber\\&\quad + \frac{\gamma(t)}{\alpha} \exp\bigg( \frac{- (x-x_L)^2}{2\sigma^2_x} + \frac{- (y-y_L)^2}{2\sigma^2_y} \nonumber \\& \quad \hspace{2cm} +\frac{- (z-z_L)^2}{2\sigma^2_z} \bigg), \quad \text{in}~~\Omega \times T. 
		\end{align}

\subsubsection{Boundary condition (BC) \eqref{eq:BC}} 
We consider the BC in \eqref{eq:BC} as a homogeneous Neumann boundary condition where we assume that the rate of the concentration of EVs at the boundary faces has no changes. It is worth noting that the assumption of a homogeneous Neumann BC for our purpose is practical because the dynamics  of the concentration of EVs degrade over time and space and would be static at the faces of 3D space for the diffusion and advection \cite{doi:10.2217/nnm.12.124}.  It is possible to consider any topology for the diffusion and advection of EVs in which the considered space has $F_i$ faces where $i\in \{1,3,4,..., N_F\}$. 

\subsubsection{Initial condition (IC) \eqref{eq:IC}} 
We consider the IC in \eqref{eq:IC} as the initial condition that determines the concentration of EVs at the initial time of the computation. To find the propagation of the released EVs from therapeutic transmitter, we assume there no EVs are present in the medium at the initial time $t_L$. 

\subsection{An Analytical Solution to the 3D-PDE} \label{sec:analytical}
We obtain an analytical solution to \eqref{eq:diff_conv_all} which enhances the reproducibility of our modeling. Considering the boundaries of  structure far from the release source, the solution is assumed for an unbounded environment. We also assume that the effect of the half-life of EVs is negligible due to their reported long half-life \cite{kwok2021extracellular}. The validity of these assumptions is verified via the numerical results presented at Section \ref{sec:results}.
 
In the 3D-PDE \eqref{eq:diff_conv_all}, $C(t)$ is obtained based on a 4D convolution of the anisotropic diffusion Green's function and the spatiotemporal EV injection function. 
For a diagonal diffusion matrix, the Green's function is given by
\begin{align}
	G\left(\mathbf{x},t\right) = G^x(x,t)G^y(y,t)G^z(z,t) \exp\left(-k_e t\right), \label{eq:green}
\end{align}
where
\begin{align}
		&G^\nu(\nu,t) \bigg|_{\nu \in \{x, y, z\}} = \dfrac{1}{\sqrt{4\pi t D_\nu}} \exp\left(-\dfrac{(\nu-v_\nu t)^2}{4D_\nu t}\right), \label{eq:green_xyz}
\end{align}
where $D_\nu \bigg|_{\nu \in \{x, y, z\}} = \dfrac{D}{\lambda^2_\nu}$. Here, the 4D convolution is implemented as 
\begin{align}\label{eq:Conv}
	&C(\mathbf{x},t) \nonumber\\&\quad=\frac{\gamma(t)}{\alpha}*\bigg[\exp\left(-k_e t\right) \left(G^x\left(x,t\right)*S^x\left(x\right)\right) \nonumber\\& \hspace{1.9cm} \times	\left(G^y\left(y,t\right)*S^y\left(y\right)\right)  
	\left(G^z\left(z,t\right)*S^z\left(z\right)\right)  \bigg],
\end{align}
where $ S^\nu\left(\nu\right) \bigg|_{\nu \in \{x, y, z\}} = \exp\left(\frac{- (\nu-\nu_L)^2}{2\sigma^2_\nu}\right)$ which is in line with \eqref{eq:release}. Eq. \eqref{eq:Conv} yields a computationally effective formulation for the spatially separable source functions. In \eqref{eq:Conv}, the first (*) at the right-hand side of the equation denotes the time convolution whereas other (*) denote the spatial convolution. To make the computation more effective, we consider \eqref{eq:Conv} in the frequency domain using the time and spatial Fourier transform given by
\begin{eqnarray}
	C(\mathbf{x},t)=\mathcal{F}^{-1}\left(\frac{\tilde{\gamma}(\omega)}{\alpha}\tilde{F}(x,y,z,\omega+k_e)\right)
\end{eqnarray}
where
\begin{subequations}
\begin{align}
	&F(\mathbf{x},t) = {F}^x(x,t){F}^y(y,t){F}^z(z,t), \\
	&{F}^x(x,t) = \mathcal{F}^{-1} \left(\tilde{G}^x(\beta_x,t)\tilde{S}^x(\beta_x,t)\right), \\
	&{F}^y(y,t) = \mathcal{F}^{-1} \left(\tilde{G}^y(\beta_y,t)\tilde{S}^y(\beta_y,t)\right),\\
	&{F}^z(z,t) = \mathcal{F}^{-1} \left(\tilde{G}^z(\beta_z,t)\tilde{S}^z(\beta_z,t)\right).
\end{align}
\end{subequations}
$\omega$ indicates the time frequency, $\beta_x,\beta_y,\beta_z$ indicate the spatial frequencies in $x,y,z$ directions, respectively, and
\begin{subequations}\label{eq:fourier}
	\begin{align}
		\tilde{G}^{\nu}\bigg|_{\nu \in \{x, y, z\}}  =\int_{-\infty}^{+\infty}G^{\nu} \exp\left(-j \nu \beta_{\nu}\right) {\text{d}\nu}, \label{eq:fourier_1}
		\\
		\tilde{S}^{\nu}\bigg|_{\nu \in \{x, y, z\}} =\int_{-\infty}^{+\infty}S^{\nu} \exp\left(-j \nu \beta_{\nu}\right) {\text{d}\nu}. \label{eq:fourier_2}
	\end{align}
\end{subequations} 

\section{Extracellular Vesicle Internalization Modeling} \label{sec:EV_Internalization}
We model the internalization of EVs through ligand-receptor interactions with the target cells and clathrin-mediated endocytosis.
 In this way, we finalize the end-to-end model for EV-based drug delivery system for cardiovascular disorders.

\subsection{Ligand-Receptor Interactions of EVs with Target Cells} \label{sec:ligand-binding}
We consider target cells as sick cardiomyocytes. We assume the target cell regenerates the binding sites at the cell membrane because of its continuous turnover, which makes us consider a constant number of binding sites as $\chi$. The dynamics of ligand-receptor interaction of EVs and the target cell is given by the following system of ODEs \cite{wilhelm2002interaction}
\begin{subequations}\label{eq:ligan-receptor}
	\begin{align} 
		\frac{\text{d} \eta_\text{b}(t)}{\text{d} t} &= \kappa_\text{a} \eta_{\text{bs}}(t) C\left(\mathbf{R_0}, t\right)  - \kappa_\text{d} \eta_\text{b}(t) - \kappa_{\text{int}} \eta_\text{b}(t), \label{eq:lig-rec-bound}
		\\
		\eta_{\text{bs}}(t) &= \chi - \eta_{\text{b}}(t), \label{eq:lig-rec-bindingsites}
		\\
		\frac{\text{d} \eta_\text{int}(t)}{\text{d} t} &= \kappa_{\text{int}} \eta_{\text{b}}(t), \label{eq:lig-rec-inter}
	\end{align} 
\end{subequations}
where $\eta_{\text{b}}(t)$, $\eta_{\text{bs}}(t)$ and $\eta_{\text{int}}(t)$ are respectively the number of bound EVs to target cell surface, the number of available binding sites, and the number of internalized EVs. Also, $\kappa_{\text{a}}$, $\kappa_{\text{d}}$ and $\kappa_{\text{int}}$ are respectively the rate of association, dissociation, and internalization of EVs. We also consider the location of target cell at $\mathbf{R_0}$ in the space, and therefore, $C\left(\mathbf{R_0}, t\right)$ is the concentration of EVs at target cell which is obtained from \eqref{eq:diff_conv_all}. To find the concentration of bound ($c^{\text{LR}}_\text{b}$) and internalized EVs ($c^{\text{LR}}_\text{int}$) by ligand-receptor interactions, we use the molar concentration rule as $c^{\text{LR}}_i\big|_{i \in \{\text{b}, \text{int} \}} = \frac{3 \eta_i}{4 \pi N_\text{A} R^3_c}$ where $R_c$ is an average radius of the target cell.

\subsection{Clathrin-Mediated Endocytosis of EVs at Target Cells} \label{sec:clathrin}

\begin{figure*}[!]
	\centering
	\subfigure[]{%
		\includegraphics[width=0.40\linewidth]{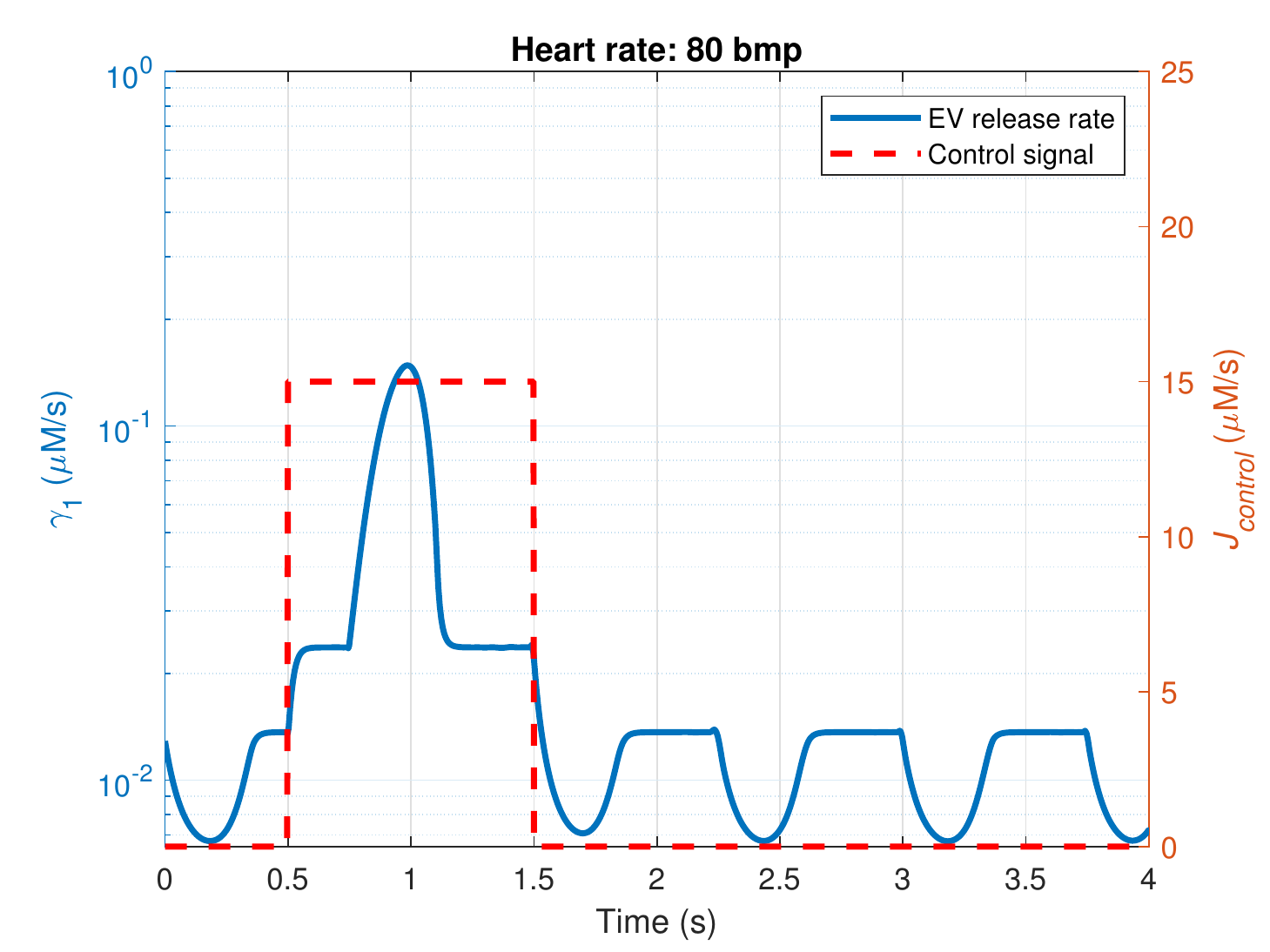}	\label{fig:release_rate_1}}%
	\subfigure[]{%
		\includegraphics[width=0.40\linewidth]{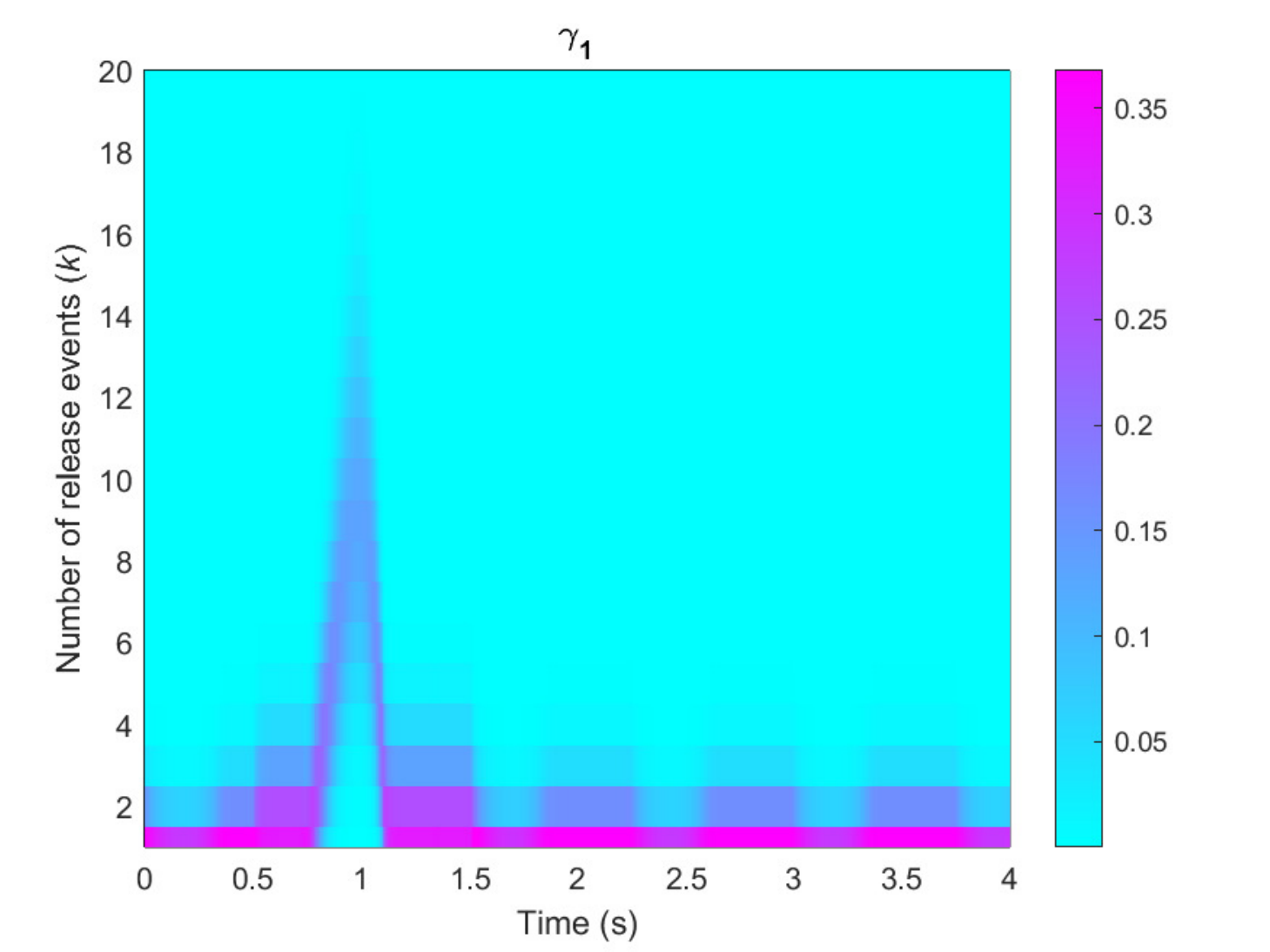}	\label{fig:release_event_1}}%
	\\
	\subfigure[]{%
		\includegraphics[width=0.40\linewidth]{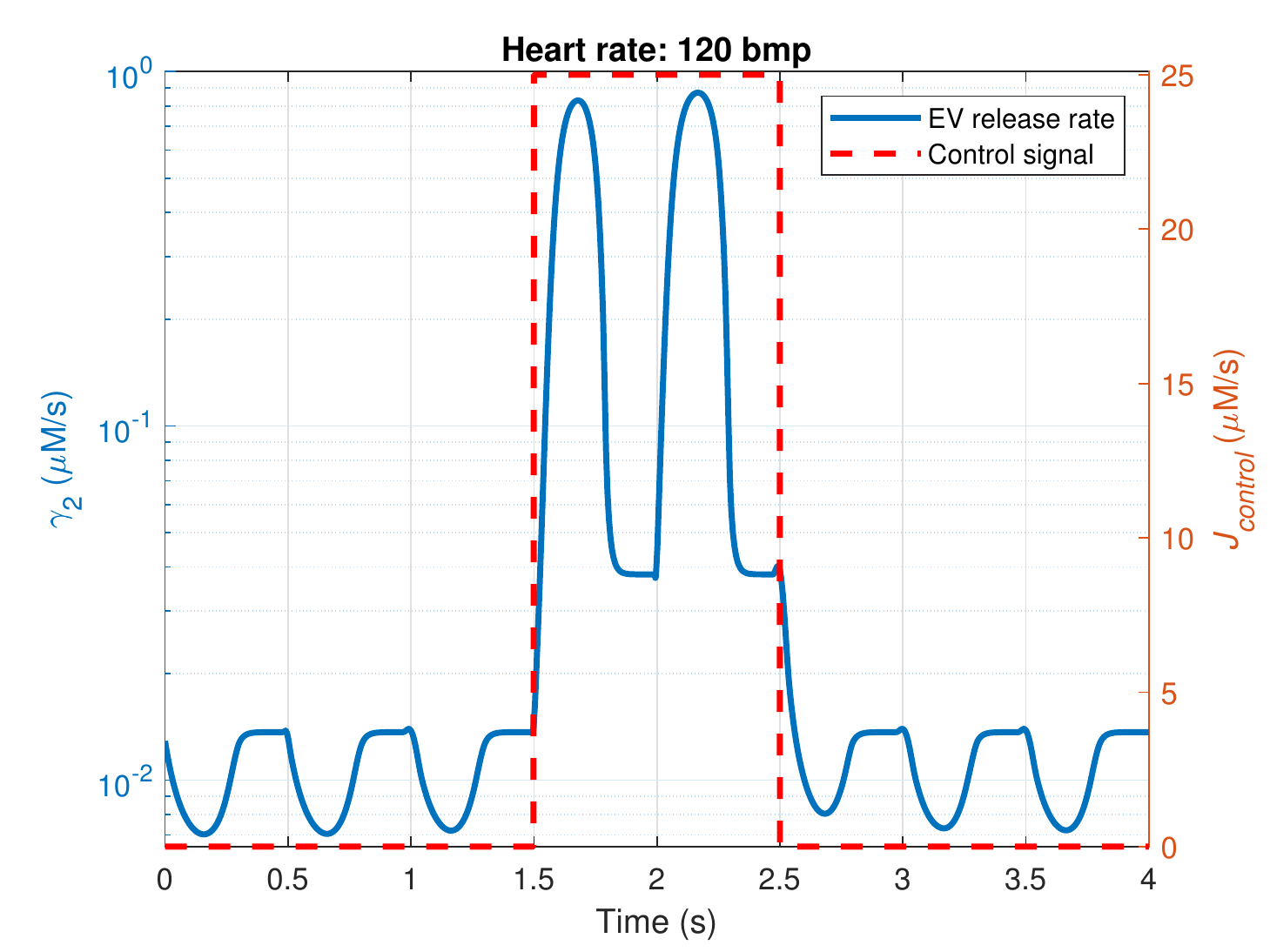}		\label{fig:release_rate_2}}%
	\subfigure[]{%
		\includegraphics[width=0.40\linewidth]{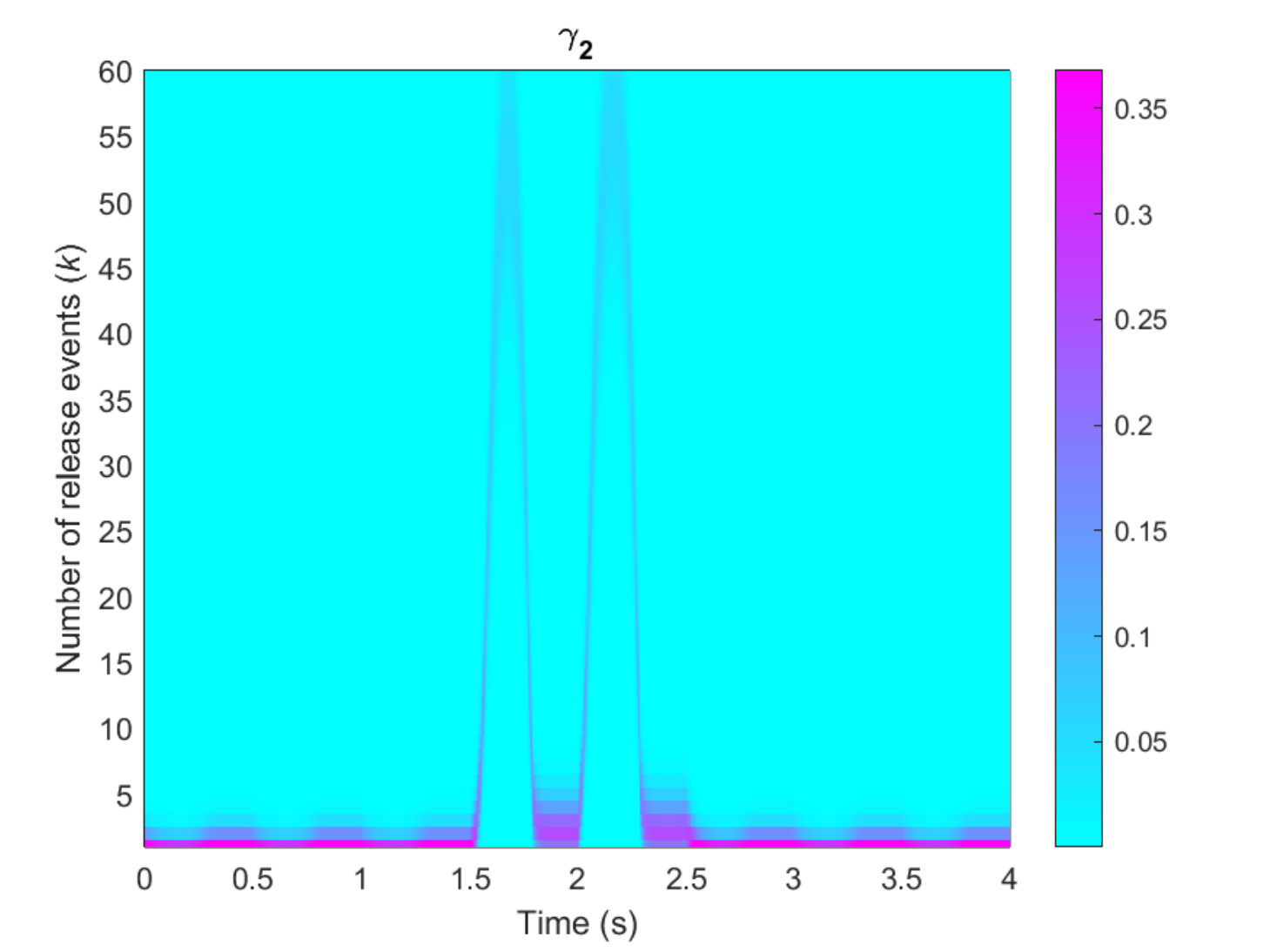}		\label{fig:release_event_2}}%
	\caption{The EV release rates are shown in (a) and (c) as functions of time for two different control signals and for two heart rates of 80 bpm and 120 bpm, respectively. The number of EV release events as a function of time according to Poisson distribution are shown in (b) and (d) based on the given EV release rates respectively shown in (a) and (c).}
	\label{fig:release_event}
\end{figure*}

The clathrin-mediated endocytosis is described by the following system of ODEs \cite{wattis2008mathematical}
\begin{subequations} \label{eq:clathrin}
	\begin{align}
		\frac{\text{d} c^{\text{CM}}_\text{b}(t)}{\text{d} t} &= a  C\left(\mathbf{R_0}, t\right) \left( p_\text{Tot} N_\text{Tot}(t) - c^{\text{CM}}_\text{b}(t) \right) - \kappa_{\text{int}} c^{\text{CM}}_\text{b}(t), \label{eq:clathrin-bound}
		\\
		\frac{\text{d} c^{\text{CM}}_\text{int}(t)}{\text{d} t} &= \kappa_{\text{int}} c^{\text{CM}}_\text{b}(t) - \kappa_\text{deg} c^{\text{CM}}_\text{int}(t), \label{eq:clathrin-inter}
	\end{align}
\end{subequations}
where $c^{\text{CM}}_\text{b}$ and $c^{\text{CM}}_\text{int}$ are respectively the concentration of bound and internalized EVs through the clathrin-mediated endocytosis. Also, $p_\text{Tot}$ and $N_\text{Tot}$ are the total number of EVs that can be coated with clathrin leading to building clathrin-coated pits, and the sum of total number of occupied and total number of unoccupied pits, respectively. In \eqref{eq:clathrin-bound}, $a = a_0/p_\text{Tot}$ where $a_0$ is the maximal binding rate of EVs. $\kappa_\text{deg}$ in \eqref{eq:clathrin-inter} is the degradation rate of EVs in the cell.

\begin{figure}[!]
	\centering
	\subfigure[]{%
		\includegraphics[width=0.50\linewidth]{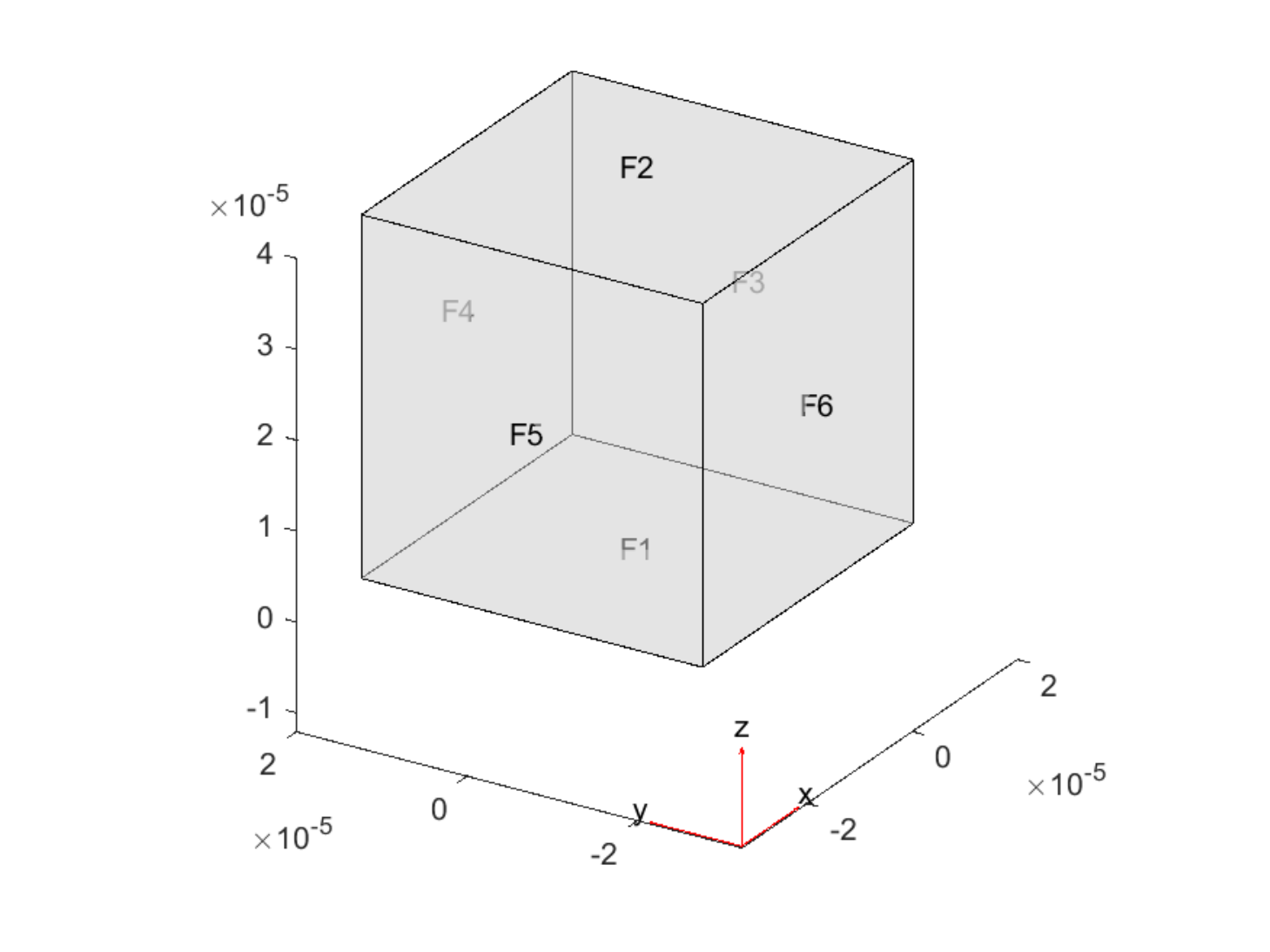}	\label{fig:model_geometry}}%
	\subfigure[]{%
		\includegraphics[width=0.50\linewidth]{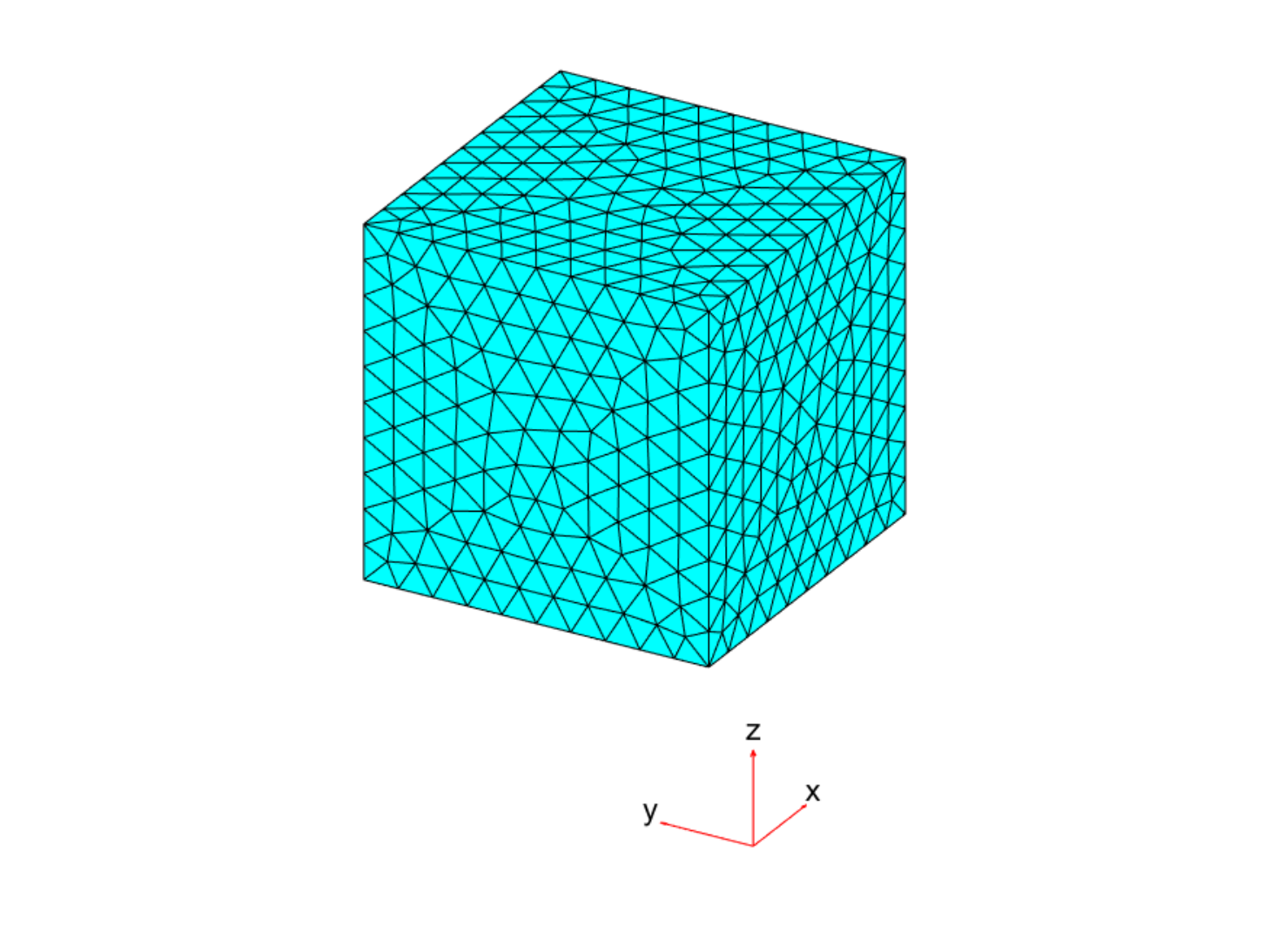}		\label{fig:model_mesh}}%
	\caption{(a) The 3D geometry model for the cardiac ECM for solving the proposed 3D-PDE. The cube in (a) comprises six 40 $\times$ 40 $\mu$m faces in a Cartesian coordinated system. (b) The mesh of the 3D geometry model given in (a) by 3D quadratic tetrahedra with a maximum edge of 4 $\mu$m.}
	\label{fig:model}
\end{figure}

 \begin{figure}[!]	
	\centering
	\includegraphics[width=\linewidth]{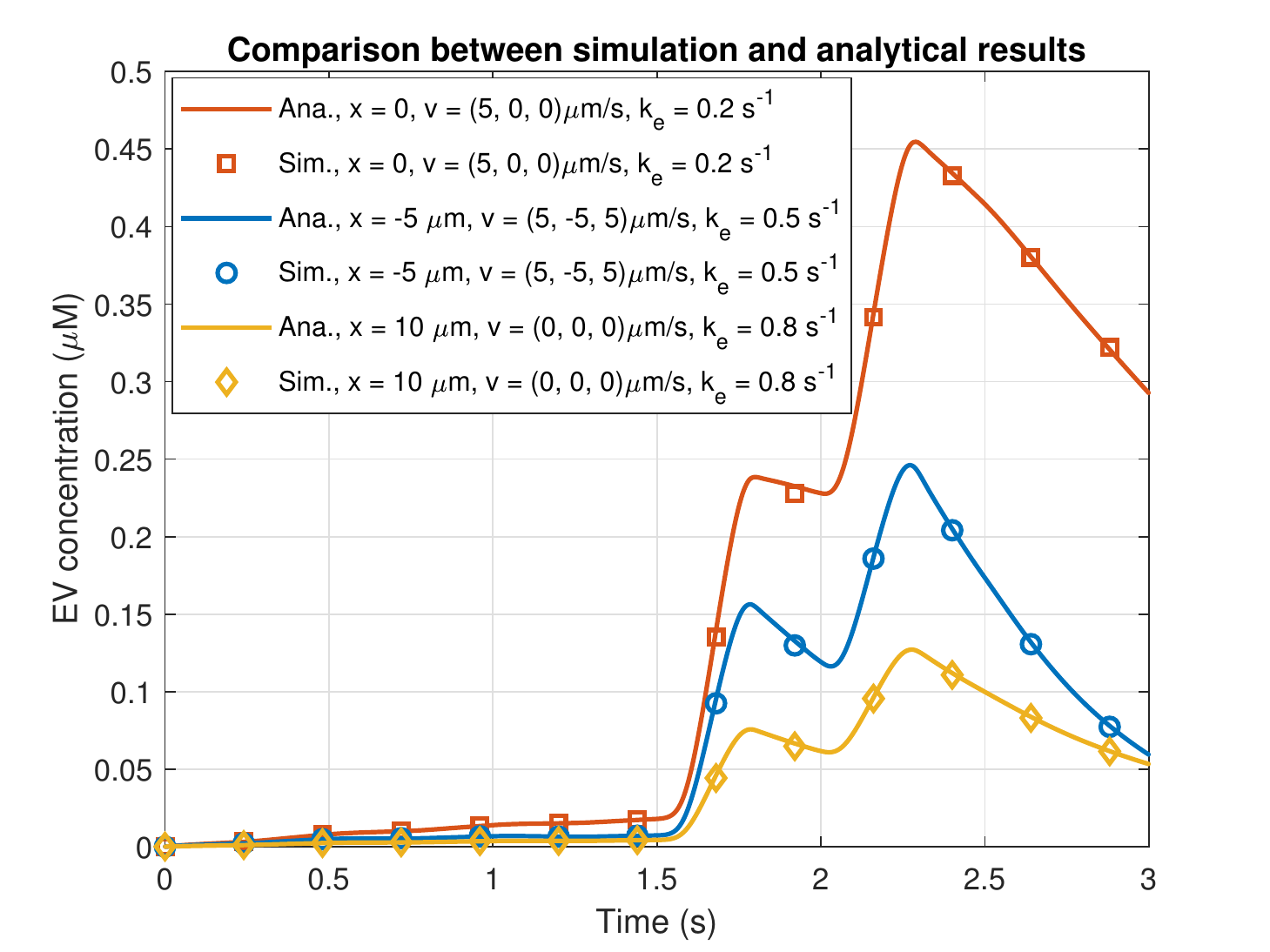}%
	\caption{The comparison between the simulation results obtained by FEM with the setting given in Fig. \ref{fig:model} and the analytical solution proposed in Section \ref{sec:analytical} is given by considering the EV concentration as a function of time at different values of $x$ while $y = 0$ and $z = 20~\mu\text{m}$ in the spatial domain.  We set the therapeutic transmitter location at $\mathbf{X_0} = \left(0, 0, 20\right) \mu\text{m}$ and the half-life of EVs as $\Lambda_{1/2} = \text{2 min}$. We compare the results based on different velocity vectors $\mathbf{v}$ and different extracellular binding degradation rates $k_e$. The analytical and simulation results show similar outcomes which verifies the assumptions made to derive the analytical solution.}\label{fig:comparison}
\end{figure}

\begin{figure*}[!]
	\centering
	\subfigure[]{%
		\includegraphics[width=0.33\linewidth]{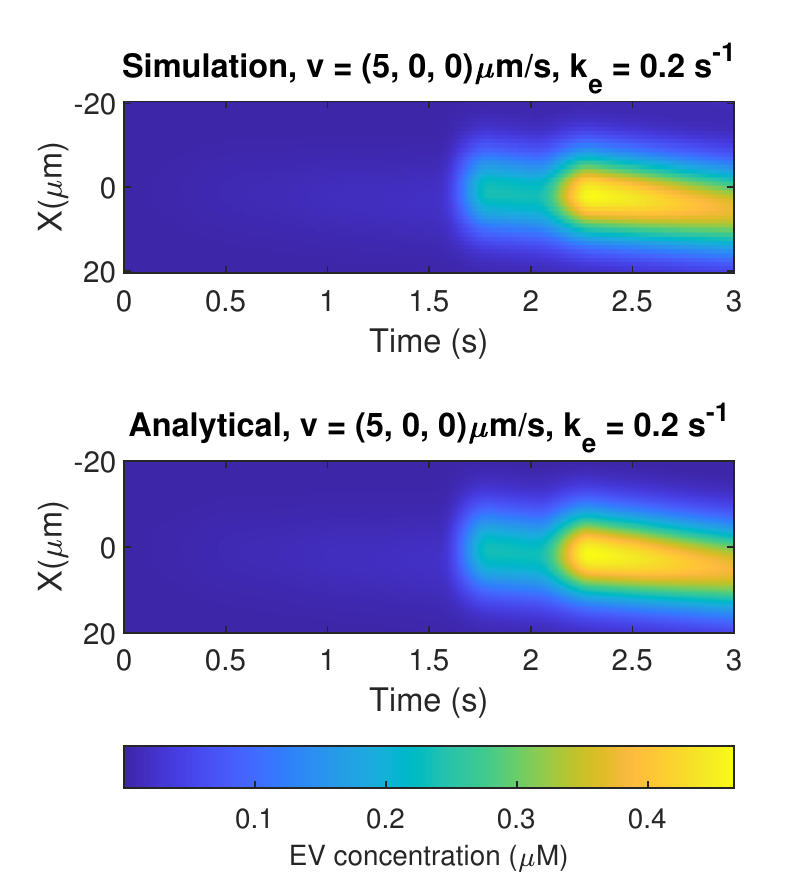}	\label{fig:EV_ECM_1}}%
	\subfigure[]{%
		\includegraphics[width=0.33\linewidth]{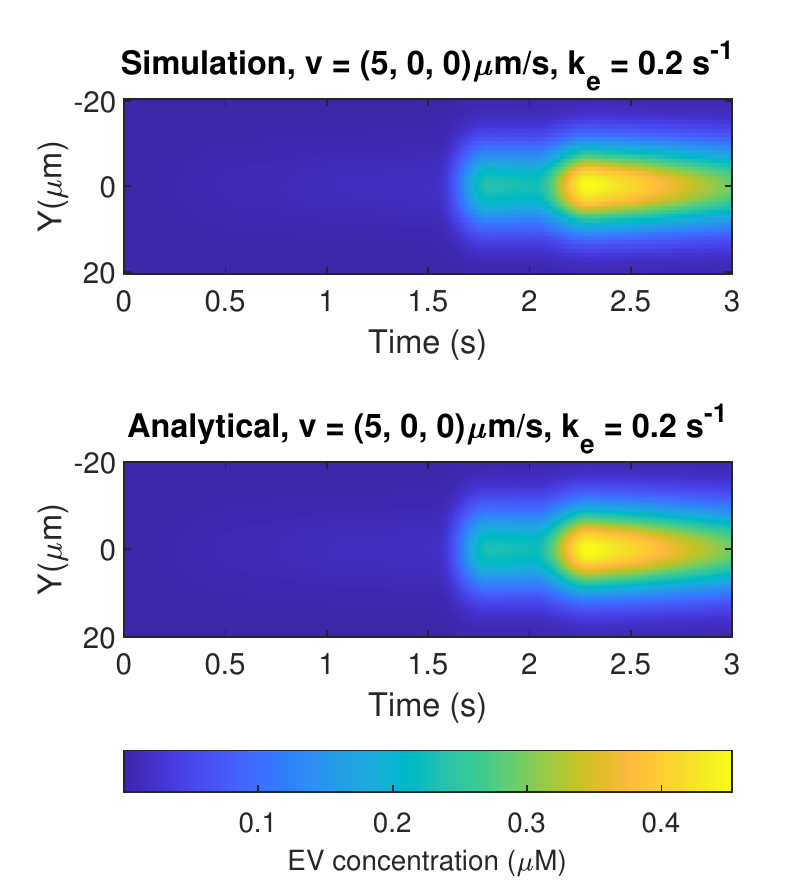}		\label{fig:EV_ECM_2}}%
	\subfigure[]{%
		\includegraphics[width=0.33\linewidth]{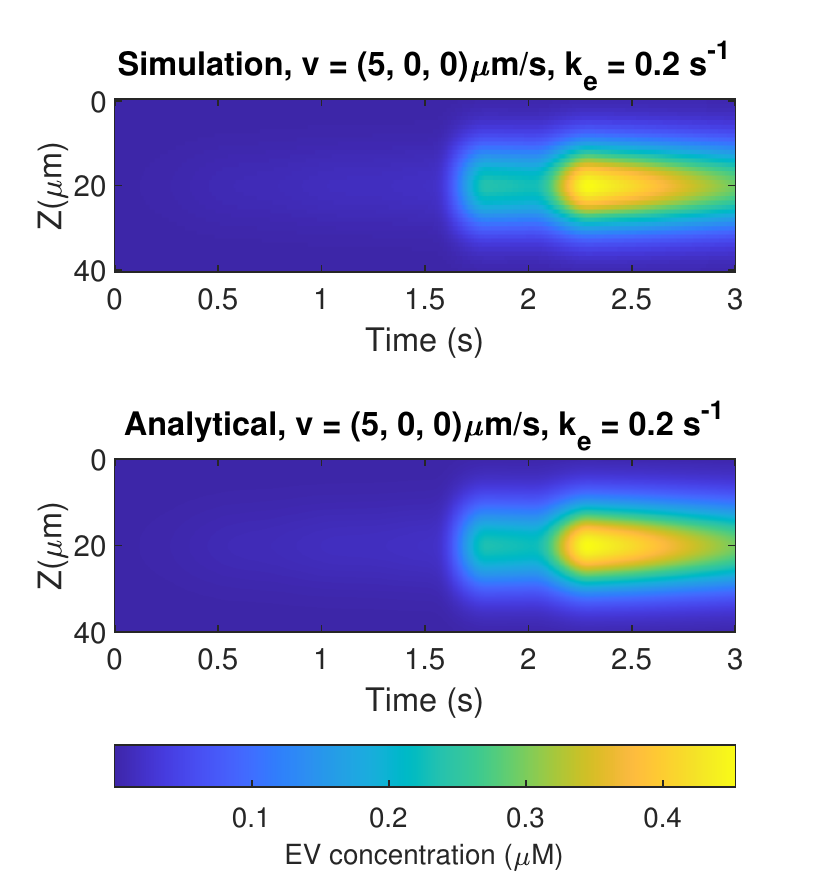}		\label{fig:EV_ECM_3}}%
	\\
	\subfigure[]{%
		\includegraphics[width=0.33\linewidth]{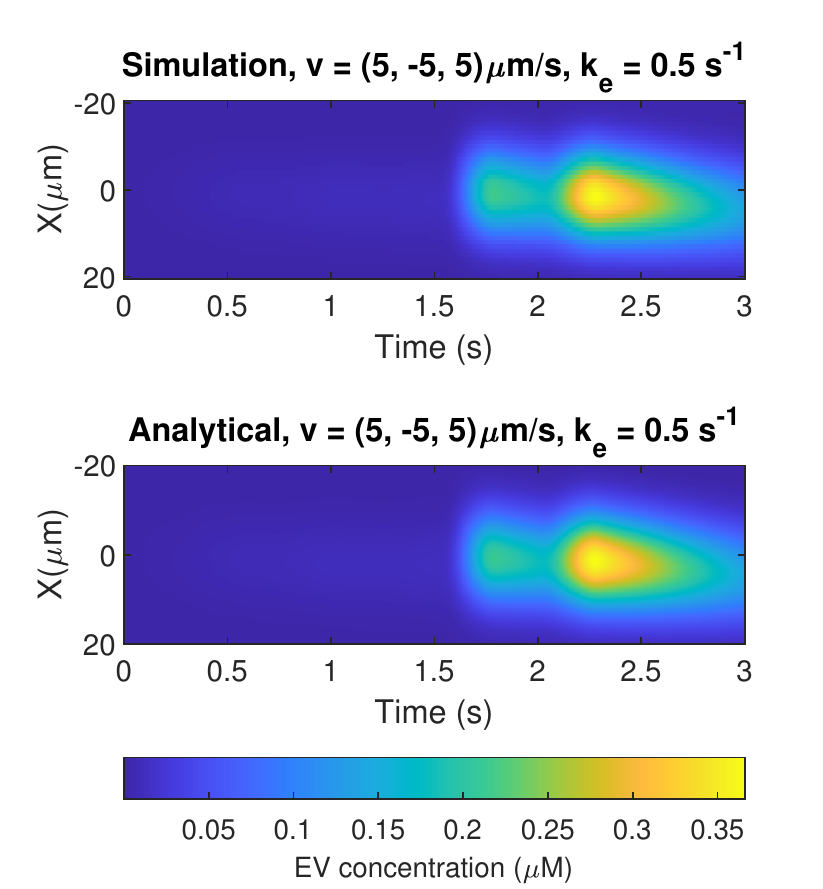}		\label{fig:EV_ECM_4}}%
	\subfigure[]{%
		\includegraphics[width=0.33\linewidth]{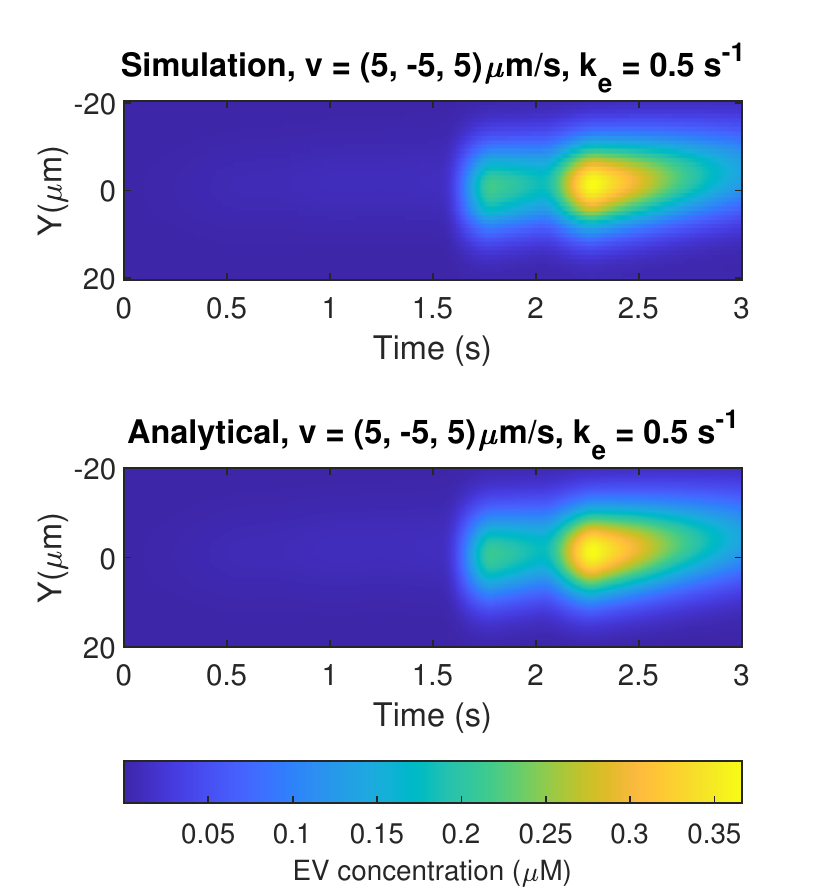}		\label{fig:EV_ECM_5}}%
	\subfigure[]{%
		\includegraphics[width=0.33\linewidth]{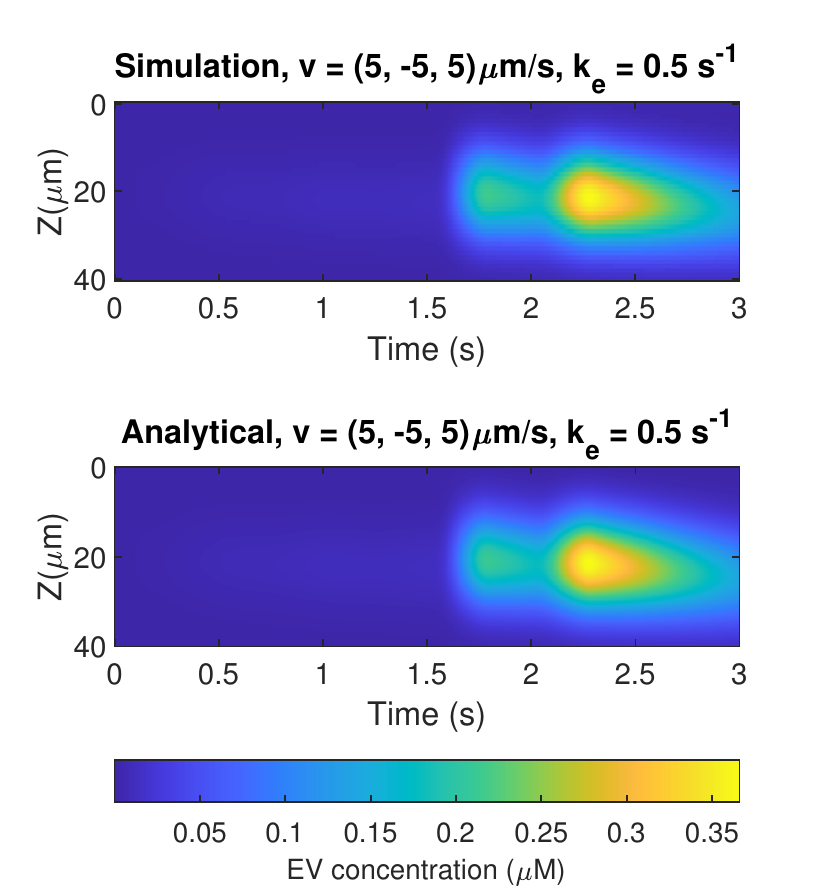} 		\label{fig:EV_ECM_6}}%
	\\
	\subfigure[]{%
		\includegraphics[width=0.33\linewidth]{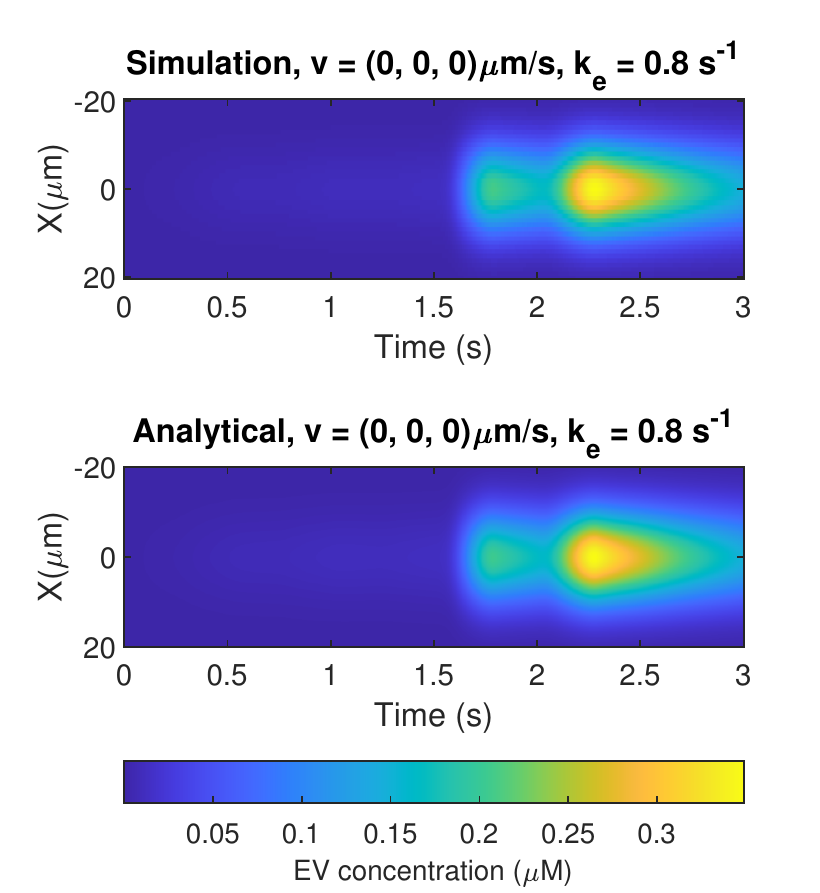} 		\label{fig:EV_ECM_7}}%
	\subfigure[]{%
		\includegraphics[width=0.33\linewidth]{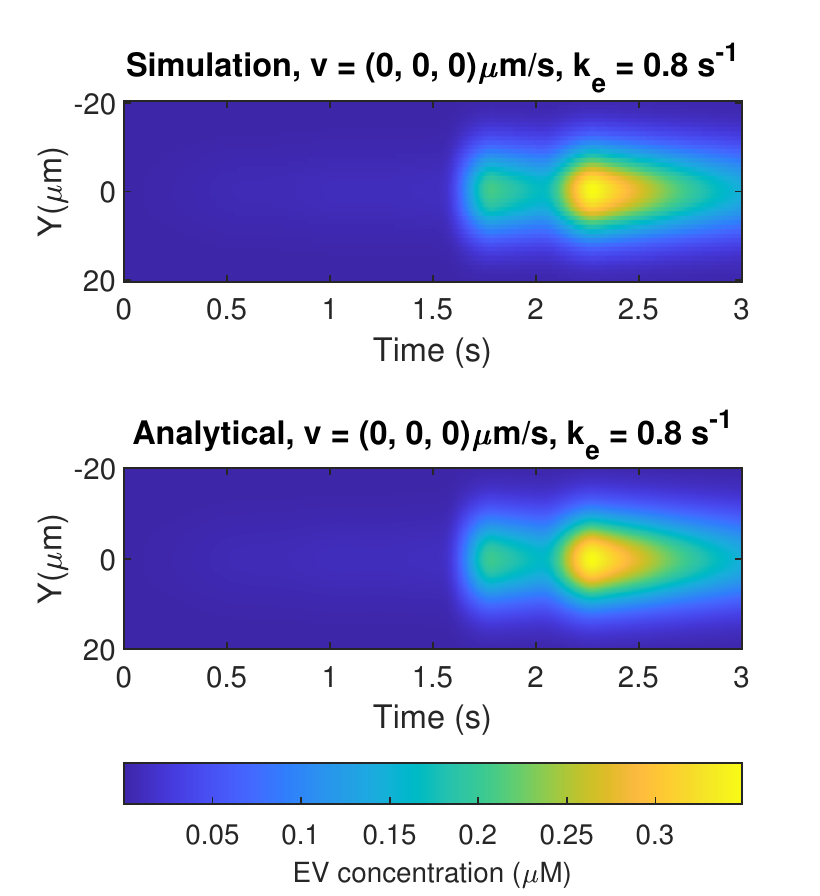} 		\label{fig:EV_ECM_8}}%
	\subfigure[]{%
		\includegraphics[width=0.33\linewidth]{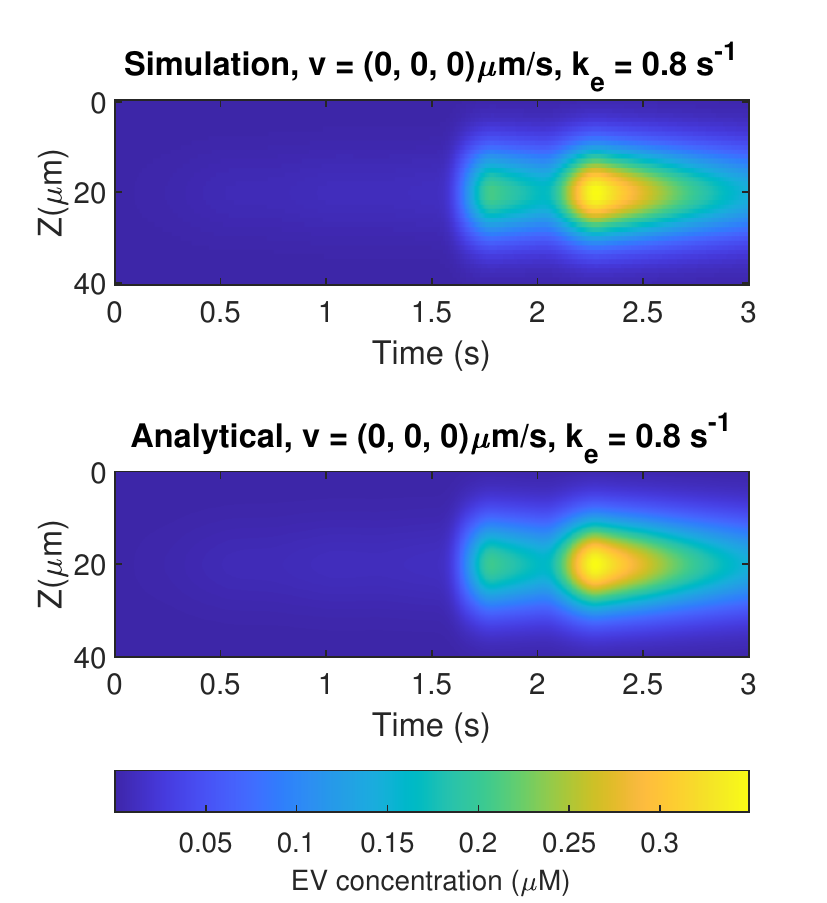} 		\label{fig:EV_ECM_9}}%
	\caption{The deflections of the concentration of EVs based on analytical and simulation results are given in $x$, $y$, and $z$ directions as a function of time, bulk flow in the ECM ($\mathbf{v}$), diffusivity tensor ($\mathbf{K}$) and the extracellular binding degradation rate ($k_e$). We set the therapeutic transmitter location at $\mathbf{X_0} = \left(0, 0, 20\right) \mu\text{m}$ and the half-life of EVs as $\Lambda_{1/2} = \text{2 min}$. (a)-(c) Scenario A: The deflections of EV concentration in the three directions by considering $\mathbf{v} = (\text{5}, \text{0}, \text{0})~\mu$m/s, $k_e = \text{0.2} \text{s}^{-\text{1}}$, and $\mathbf{K} = \begin{bmatrix}D/\text{1.1}^\text{2} & D/\text{1.4}^\text{2} & D/\text{1.7}^\text{2}\end{bmatrix} \times \mathbf{I}$. (d)-(f) Scenario B: The deflections of EV concentration in the three directions by considering $\mathbf{v} = (\text{5}, \text{-5}, \text{5})~\mu$m/s, $k_e = \text{0.5} \text{s}^{-\text{1}}$, and $\mathbf{K} = \begin{bmatrix}D/\text{1.1}^\text{2} & D/\text{1.1}^\text{2} & D/\text{1.1}^\text{2}\end{bmatrix} \times \mathbf{I}$. (g)-(i) Scenario C: The deflections of EV concentration in the three directions by considering $\mathbf{v} = (\text{0}, \text{0}, \text{0})~\mu$m/s, $k_e = \text{0.8} \text{s}^{-\text{1}}$, and $\mathbf{K} = \begin{bmatrix}D/\text{1.4}^\text{2} & D/\text{1.4}^\text{2} & D/\text{1.4}^\text{2}\end{bmatrix} \times \mathbf{I}$.}
	\label{fig:EV_ECM}
\end{figure*}

\section{Numerical Results and Discussion} \label{sec:results}
This section presents the numerical results of the proposed end-to-end EV-based drug delivery system by studying the EV release process, propagation, and internalization. We mainly use the parameters from \cite{9488662} for the simulation of the EV release process. Other values relevant for the simulations and reproducibility of the results are given in Appendix ~\ref{app:parameters}.

Fig. \ref{fig:release_event} shows the EV release rates and processes modulated by external control device which induces current in the EV releasing cell. Fig \ref{fig:release_rate_1} shows the EV release rate by considering the heart rate as 80 beat per minute (bpm) while the external device depolarizes the cell and affects Ca$^\text{2+}$ dynamics by a pulse signal with an amplitude of 15 $\mu$M/s and 1 s duration. This accordingly modulates the cumulative EV release rate around 0.1 $\mu$M/s in the given time window. We consider an average of 24 EVs in each MVB \cite{von2011multivesicular} and $\Delta t =$ 5 ms. The number of release events ($k$) follows a Poisson distribution in each time interval of $(t, t+\Delta t]$ as shown in Fig. \ref{fig:release_event_1}. According to Fig \ref{fig:release_event_1}, the therapeutic transmitters presumably have a maximum number of 15 release events in each time interval in the time frame of modulated Ca$^\text{2+}$ levels. With a greater amplitude of control signal of 25 $\mu$M/s, Fig. \ref{fig:release_rate_2} shows that the therapeutic transmitters can have the EV release rate of 1 $\mu$M/s. The heart rate in Fig. \ref{fig:release_rate_2} is 120 bpm which shows that the frequency of release rate is higher than in Fig. \ref{fig:release_rate_1}. According to Fig. \ref{fig:release_event_2}, the therapeutic transmitters presumably have a maximum of 60 release events in each $(t, t+\Delta t]$ time interval. Thereby, we conclude that a higher amplitude of control signal leads to a greater number of release events by considering a Poisson process for the EV release. We also infer that the maximum number of release events most likely happens concurrently with the EV release rate's peaks.

We present a numerical simulation of the 3D-PDE proposed in \eqref{eq:diff_conv_all} using the Partial Differential Equation Toolbox in \textsc{Matlab}~\cite{matlabpdet}. We first create a 3D geometry of a cube displayed in Fig. \ref{fig:model_geometry} with six 40 $\times$ 40 $\mu$m square faces and its center located in $(\text{0}, \text{0}, \text{20})~\mu$m. The mesh elements in the geometry of the cardiac ECM are 3D quadratic tetrahedra with a maximum edge length of 4 $\mu$m as shown in Fig. \ref{fig:model_mesh}. We can reach a more detailed solution by decreasing the maximum edge length of the mesh elements; however, the simulation time increases greatly. Nevertheless, we select the maximum mesh size which gives a balance between the approximate solution and simulation time and enhances the reproducibility of this modeling. The PDE Toolbox in \textsc{Matlab} solves the PDEs using FEM; to understand the underlying principle of this numerical method, we give a basic introduction in Appendix~\ref{appendix:FEM} to solve the proposed 3D-PDE using FEM.

In Fig. \ref{fig:comparison}, we compare the simulation results obtained by FEM and the analytical solution given in Section \ref{sec:analytical}. The setting for the simulation results is the same as the setting displayed in Fig. \ref{fig:model}. We consider the unbounded ECM for studying the analytical solution. We set the therapeutic transmitters at $\mathbf{X_0} = \left(0, 0, 20\right) \mu\text{m}$ and show the simulation and analytical results for the EV concentration as a function of time by considering different velocity vectors and degradation rates $k_e$. Also, we consider the EV source as $\gamma_2(t)$ which is given in Fig. \ref{fig:release_rate_2} when the heart rate is 120 bpm and the control signal has an amplitude of 25 $\mu \text{M}/\text{s}$. Fig. \ref{fig:comparison} shows the EV concentration with different $x$ as a function of time when $y = 0$ and $z = \text{20}~\mu\text{m}$. The EV concentration for closer view points is higher than for far located view points. Fig. \ref{fig:comparison} demonstrates that the increase in the extracellular binding degradation rate $k_e$ decreases the EV concentration at different locations. Fig. \ref{fig:comparison} also demonstrates that the analytical solution correctly predicts the EV concentration considering $\gamma_2(t)$ as the EV release rate and the EV concentration peaks follow the EV release rate's peaks in the time period of $[\text{1.5}~\text{2.5}]$s. It is worth noting that the analytical solution can successfully predict the EV concentration using any type of the injection model (not only a Gaussian function as used in the presented analysis).  Fig. \ref{fig:comparison} thus verifies the assumption of having an unbounded environment for the analytical solution when the therapeutic transmitters are located far from the boundaries of the structure given in Fig. \ref{fig:model_geometry}. Also, Fig. \ref{fig:comparison} verifies the assumption that the half-life of EVs have a negligible effect on their propagation. 

We present the EV concentration dynamics in Fig.~\ref{fig:EV_ECM} with deflections shown in each dimension of the Cartesian coordinate system for both FEM simulation and analytical solution. We consider the model geometry given in Fig.~\ref{fig:model} and the EV release rate $\gamma_2$ given in Fig. \ref{fig:release_rate_2}. We place the therapeutic transmitters at point $\mathbf{X_0} = \left(0, 0, 20\right) \mu\text{m}$ and $\Lambda_{1/2} = \text{2 min}$, considering three scenarios based on different values of the velocity vector $\mathbf{v}$, extracellular binding degradation rate $k_e$, and diffusivity tensor $\mathbf{K}$, as follows
\begin{itemize}
	\item Scenario A: $\mathbf{v} = (\text{5}, \text{0}, \text{0})~\mu$m/s, $k_e = \text{0.2} \text{s}^{-\text{1}}$, $\mathbf{K} = \begin{bmatrix}D/\text{1.1}^\text{2} & D/\text{1.4}^\text{2} & D/\text{1.7}^\text{2}\end{bmatrix} \times \mathbf{I}$,
	\item Scenario B: $\mathbf{v} = (\text{5}, \text{-5}, \text{5})~\mu$m/s, $k_e = \text{0.5} \text{s}^{-\text{1}}$, $\mathbf{K} = \begin{bmatrix}D/\text{1.1}^\text{2} & D/\text{1.1}^\text{2} & D/\text{1.1}^\text{2}\end{bmatrix} \times \mathbf{I}$,
	\item Scenario C: $\mathbf{v} = (\text{0}, \text{0}, \text{0})~\mu$m/s, $k_e = \text{0.8} \text{s}^{-\text{1}}$, $\mathbf{K} = \begin{bmatrix}D/\text{1.4}^\text{2} & D/\text{1.4}^\text{2} & D/\text{1.4}^\text{2}\end{bmatrix} \times \mathbf{I}$.
\end{itemize}



Figs. \ref{fig:EV_ECM_1}-\ref{fig:EV_ECM_3} show the results of Scenario A. In this scenario, there is a diagonal shape at the peak of the EV concentration which is because of the velocity in the $x$ direction. There are two peaks in the EV concentration because of the EV release rate $\gamma_2$. Figs.~\ref{fig:EV_ECM_2} and \ref{fig:EV_ECM_3} show the deflections of the EV concentration respectively in $y$ and $z$ directions. The concentrations' peaks and overall values of the EV concentration in $y$ and $z$ directions are similar in Scenario A because of the zero bulk flow in these directions, and the symmetrical transmitters' location for these directions compared to the geometry model. Another reason for similar EV concentration deflections in $y$ and $z$ directions is the equal values of $\sigma_x$, $\sigma_y$, and $\sigma_z$, making a 3D Gaussian form of concentration around the center of the cube. The distribution of the EV concentration in Fig. \ref{fig:EV_ECM_1} is also different from Figs. \ref{fig:EV_ECM_2} and \ref{fig:EV_ECM_3} because the bulk flow in the $x$ direction is higher than in other directions, which results in higher concentrations of EVs. Also, although the tortuosity of the ECM is different in each direction in Scenario A, Figs. \ref{fig:EV_ECM_2} and \ref{fig:EV_ECM_3} indicate that it has less impact on the overall EV concentration. It is because of the small value of the diffusion coefficient of EVs in the cardiac ECM; however, the anisotropic behavior of the ECM is considered in the modeling and can be assessed through required setting. 
 
 
 \begin{figure}[!]
 	\centering
 	\includegraphics[width=\linewidth]{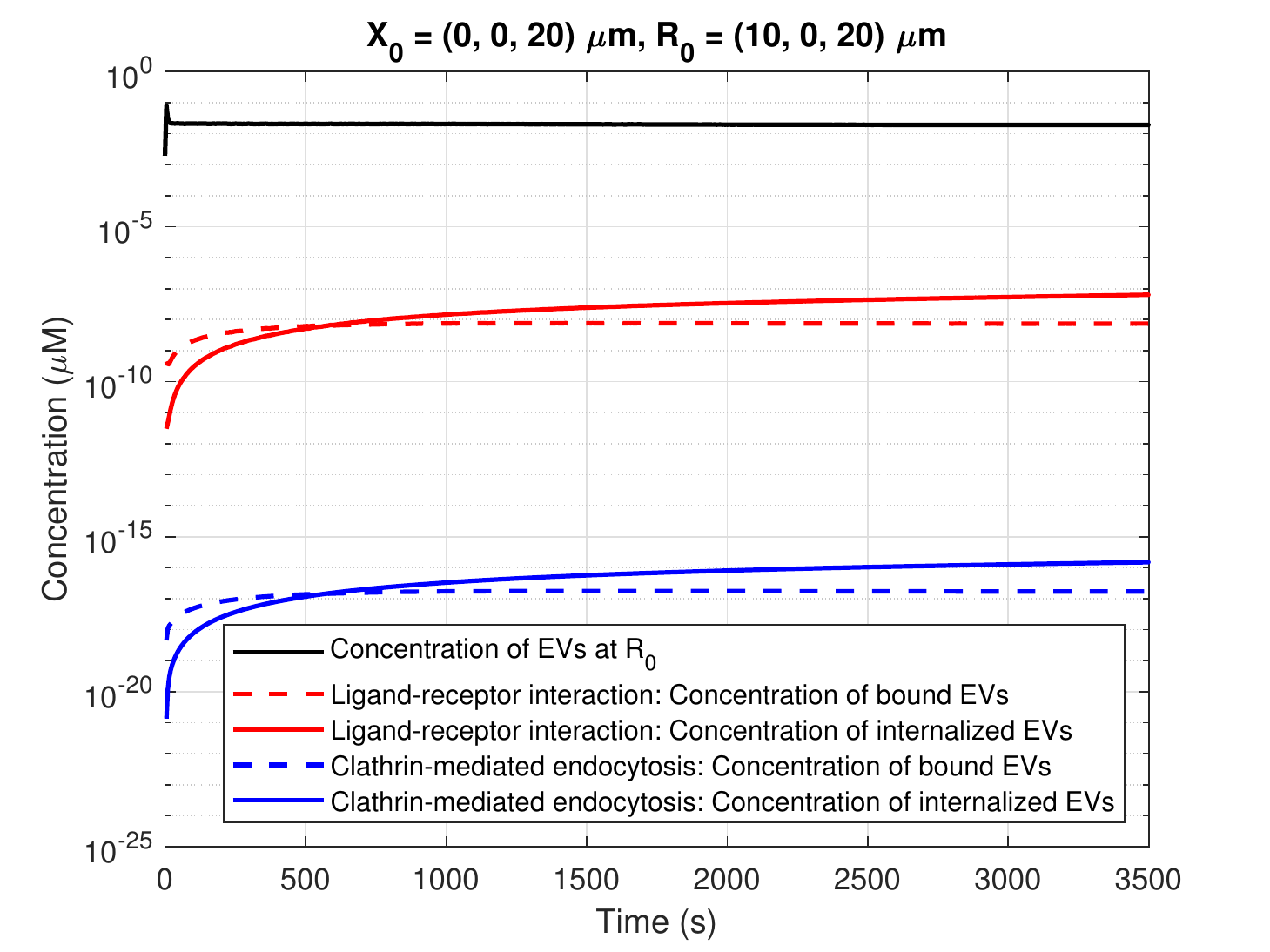}%
 	\caption{The concentration of bound and internalized EVs by considering two internalization methods, i.e., ligand-receptor interaction and clathrin-mediated endocytosis as functions of time when the therapeutic transmitters are located at $\mathbf{X_0} = (\text{0}, \text{0}, \text{20})~\mu$m and the target cells are located at $\mathbf{R_0} = (\text{10}, \text{0}, \text{20})~\mu$m.}
 	\label{fig:receiver}
 \end{figure}

The results of Scenario B are shown in Figs. \ref{fig:EV_ECM_4}-\ref{fig:EV_ECM_6}. 
There are diagonal shapes for the EV concentration in $x$, $y$, and $z$ directions because of the bulk flow in these directions. The deflections of the EV concentration in the three directions in Scenario B are less than Scenario A because of the higher extracellular binding degradation rate. Finally, the results of Scenario C are shown in Figs. \ref{fig:EV_ECM_7}-\ref{fig:EV_ECM_9}. In this scenario, there is no bulk flow in any direction and hence there is no diagonal shape of the EV concentration in Figs. \ref{fig:EV_ECM_7}-\ref{fig:EV_ECM_9}. The EV concentration in Scenario C is smaller than other scenarios which stems from the higher extracellular binding degradation rate. We also present an animation showing the EV concentration in the cardiac ECM in Fig. S1 as the supplementary file. The simulation for the animation is created using COMSOL Multiphysics.

Fig. \ref{fig:EV_ECM} provides preliminary results for further testing of the cardiac drug delivery system. The geometry model for the ECM can also be enhanced considering more complex and practical geometries derived using CT scan imaging. In this regard, the CT scan image can be converted to a geometry model and imported to the PDE proposed in \ref{eq:diff_conv_all}. The proposed ECM modeling and propagation of EVs can be used with any geometry and release scheme models; nevertheless, the main 3D-PDE and BC need to be revised based on the application and experimental inputs.

We present the simulation results of the EV internalization  in Fig. \ref{fig:receiver} for two internalization mechanisms, i.e., ligand-receptor interactions and clathrin-mediated endocytosis, when the therapeutic transmitters are located at  $\mathbf{X_0} = (\text{0}, \text{0}, \text{20})~\mu$m and target cells are located at  $\mathbf{R_0} = (\text{10}, \text{0}, \text{20})~\mu$m. As Fig. \ref{fig:receiver} shows, the internalization of EVs at the target cells takes long time and leads to the maximum level of $\text{10}^{\text{-7}} \mu$M EVs after 3 $\times \text{10}^{\text{3}}$ s. Fig. \ref{fig:receiver} also demonstrates that the target cells can internalize more EVs through ligand-receptor interactions rather than clathrin-mediated endocytosis. Fig.~\ref{fig:receiver} also reveals that the concentration of bound EVs via both internalization methods is higher than the concentration of internalized EVs until a specific time around 500 s. This implies that the concentration of bound EVs should reach a threshold until the EVs internalize with higher concentration at target cells. 

\section{Conclusion} \label{sec:conclusion}
Mathematical modeling helps scientists test and develop novel treatment approaches for different types of disorders. Here, we have modeled an end-to-end drug delivery system based on extracellular vesicles (EVs), to treat cardiovascular diseases (CVDs). The results from our analyzed end-to-end drug delivery system can be applied to other carriers such as liposomes and utilized for treatment. We have utilized the molecular communication (MC) theory as the methodology for the EV release, propagation and internalization modeling. We have modeled the stochastic nature of the EV release from human induced pluripotent stem cell (HiPSC)-derived cardiomyocytes at the release part using the Poisson process. Our findings show a strong correlation between the number of the EV release events and the control signal peaks. We have modeled the propagation of EVs through the cardiac extracellular matrix (ECM) using 3-dimensional (3D) partial differential equations (PDEs), and obtained an analytical solution to the 3D-PDE by a Green's function. The analytical solution is verified through finite element method simulations. Our findings reveal that the EVs' concentration dynamics depends on the modeling parameters of the cardiac ECM, such as volume fraction and tortuosity. In more convoluted pathways, EVs diffuse more slowly; however, bulk flow in the ECM can mediate EVs to reach distant target cells with higher concentrations. Ultimately, we have modeled the internalization of EVs based on two methods (ligand-receptor interactions and clathrin-mediated endocytosis) by systems of ordinary differential equations (ODEs). By comparing the two internalization methods, our findings show that although the internalization process at target cells slowly occurs, ligand-receptor interactions can lead to more internalized EVs rather than clathrin-mediated endocytosis. Furthermore, our results indicate that when the concentration of bound EVs reaches a threshold, the concentration of internalized EVs increases over the bound EVs. The proposed mathematical modeling of the EV-mediated end-to-end drug delivery can be used for examining novel treatment approaches of CVDs and potentially other types of disorders in future health applications. 

Although our modeling provide preliminary results, it can be further developed by importing computational tomography (CT) scan imaging methods to the geometry model of the cardiac ECM for more accurately designed EV-mediated delivery systems of heart applications. Also, biological systems have naturally complex structures, which can affect the overall results. Hence, the proposed mathematical modeling needs to be verified through experimental studies on cells and further on animal experiments. We aim to use experimental studies to validate or modify our modeling results. Furthermore, the contraction of the heart will modify the shape and characteristics of the ECM which will require further sophistication of the modeling.


\appendices

\section{Values and Parameters for the Simulations} \label{app:parameters}
We list the parameters for the simulations in~Table~\ref{table}.

\begin{table}[!]
	\caption{Values and Parameters for the Simulations of End-to-End EV-based Drug Delivery}
	\begin{center}
		\begin{tabular}{c|c|c}
			Parameter& Value & Reference \\ \hline
			$\mathbb{E}[N_\text{MVB}]$& 24 &\cite{von2011multivesicular} \\
			$N_\text{A}$ &  6.02214086 $\times$10$^\text{23}$ mol$^\text{-1}$\\
			$d_\text{MVB}$ & 500 nm &\cite{von2011multivesicular}\\
			$\Delta t$ & 5 ms&\\
			$\lambda_x$, $\lambda_y$, $\lambda_z$& $\{\text{1.1}, \text{1.4}, \text{1.7}\}$ &\cite{lenzini2020matrix}\\
			$D$ & 1 $\mu$m$^\text{2}$/s & \cite{lenzini2020matrix}\\
			$\alpha$ & 0.6 & \cite{miller2013comprehensive}\\
			$\Lambda_{1/2}$ & 2 min &\cite{kwok2021extracellular}\\
			$k_e$ & [0.0008, 0.8] s$^{-\text{1}}$& \cite{sung2015directional}\\
			$\sigma_x$, $\sigma_y$, $\sigma_z$ & 7 $\times$ 10$^\text{-6}$&\\
			$t_L$ & 0& \\
			$t_R$ & [3, 5000] s &\\
			$\kappa_{\text{a}}$ & 10$^\text{4}$ M$^\text{-1}$ s$^\text{-1}$ &\cite{chiodi2021multiplexed,wilhelm2002interaction}\\
			$\kappa_{\text{d}}$ & 10$^\text{-10}$ s$^\text{-1}$ &\cite{chiodi2021multiplexed}\\
			$\kappa_{\text{int}}$ & 0.0027 s$^\text{-1}$& \cite{wattis2008mathematical}\\
			$\kappa_\text{deg}$ & 0.0002 s$^\text{-1}$ &\cite{wattis2008mathematical}\\
			$\chi$ & 5.3 $\times$ 10$^\text{4}$ &\cite{goodman2008spatio}\\
			$R_c$ & 82.5 $\mu$m &\cite{VU2014127}\\
			$a$ & 6.64 $\times$ 10$^\text{-17}$ &\cite{wattis2008mathematical}\\
			$p_\text{Tot}$ & 200 &\cite{wattis2008mathematical}\\
			$N_\text{Tot}$ & 180 &\cite{wattis2008mathematical}\\		
		\end{tabular} \label{table}
	\end{center}
\end{table}

\section{An Introduction to Finite Element Method Solution of The 3D-PDE} \label{appendix:FEM}
We study a basic introduction to FEM to understand the underlying principles of numerical analysis for the computer simulations of the propagation of EVs in the ECM. FEM is a numerical solution to differential equations such as PDEs based on a subdivision of large space variables into smaller and simpler parts called finite elements. The subdivision is achieved by discretization of space using mesh construction of the object \cite{reddy2019introduction}. 

To numerically solve the advection-diffusion 3D-PDE, we first convert the strong form of the 3D-PDE problem of \eqref{eq:diff_conv_all} to a weak form by multiplying the equation with a test function denoted by $\vartheta$ and integrating over the spatial domain as
\begin{align} \label{eq:FEM_1}
	&\int_{\Omega}  \vartheta \frac{\partial C\left(\mathbf{x},t\right)}{\partial t} \text{d}\Omega  + \int_{\Omega}  \left( \mathbf{K} \vec{\nabla} \vartheta  - \mathbf{v} \right) \vec{\nabla} C\left(\mathbf{x},t\right)  \text{d}\Omega \nonumber\\&\quad- \int_{\partial\Omega_N} \vec{n}\cdot\left(\textbf{K} \vec{\nabla} C\left(\mathbf{x},t\right)\right) \text{d}\Omega_N \nonumber \\&\quad  = \int_{\Omega} \vartheta \left(P(t) - \Gamma(\mathbf{x}, t, \mathbf{X_0})\right) \text{d}\Omega, \hspace{0.75cm}\forall \vartheta.
\end{align}
The Neumann BC in \eqref{eq:BC} nullifies the last term from the left-hand side of \eqref{eq:FEM_1}. Next, we discretize the weak form in \eqref{eq:FEM_1} by subdividing the spatial domain into smaller subdomains of $\Omega^e$ where $\Omega = \cup \Omega^e$. Then, we represent the finite-dimensional equivalent of admissible and trial functions respectively denoted by $c_h$ and $\vartheta_h$ where the discretized form of the weak function is 
\begin{align} \label{eq:FEM_2}
	&\int_{\Omega^e}  \vartheta_h \frac{\partial c_h\left(\mathbf{x},t\right)}{\partial t} \text{d}\Omega^e  + \int_{\Omega^e}  \left( \mathbf{K} \vec{\nabla} \vartheta_h  - \mathbf{v} \right) \vec{\nabla} c_h\left(\mathbf{x},t\right)  \text{d}\Omega^e \nonumber \\&\quad  = \int_{\Omega^e} \vartheta_h \left(P(t) - \Gamma(\mathbf{x}, t, \mathbf{X_0})\right) \text{d}\Omega^e, \hspace{0.75cm}\forall \vartheta_h.
\end{align}
Finally, we utilize piece-wise polynomial basis functions of $\varTheta_j$ where $j \in \{1, 2, ..., N_c\}$ so that any approximated solution $c_h$, can be considered as a linear combination of basis functions as
\begin{align} \label{eq:FEM_3}
	c_h\left(\mathbf{x},t\right) = \sum_{j=1}^{N_c} c_j\left(t\right) \varTheta_j(\mathbf{x}).
\end{align}
Hence, we have a system of $N_c$ ODEs to solve where $c_j$ functions are undetermined. FEM makes an approximation to the solution by minimizing an associated error function.

\bibliographystyle{ieeetr}
\bibliography{Hamid}

\end{document}